\documentclass{emulateapj}

\newcommand {\ergs} {erg s$^{-1}$}
\newcommand {\ergcms} {erg cm$^{-2}$ s$^{-1}$}
\received{2002 October 1}
\slugcomment{Accepted for August 2003 ApJ}
\begin{document}
\title{A Chandra Survey of the Nearest ULIRGs: Obscured AGN or
  Super-Starbursts?}
\author{A. Ptak, T. Heckman}
\affil{The Johns Hopkins University, Dept. of Physics and Astronomy, 
Baltimore, MD 21218}
\author{N. A. Levenson}
\affil{Department of Physics and Astronomy, University of Kentucky,
Lexington, KY 40506} 
\author{K. Weaver\altaffilmark{1}}
\affil{NASA Goddard Space Flight Center, Laboratory for High Energy
  Astrophysics} 
\author{D. Strickland\altaffilmark{2}}
\affil{The Johns Hopkins University, Dept. of Physics and Astronomy, 
Baltimore, MD 21218}
\altaffiltext{1}{Also at Johns Hopkins University}
\altaffiltext{2}{Chandra Fellow}
\begin{abstract}
% --- redone to be 250 words
We present initial results from a Chandra survey of a complete sample
of the 8 nearest ($z \leq$ 0.04) ultraluminous IR galaxies
(ULIRGs), and also include the IR-luminous galaxy NGC 6240 for comparison. 
In this paper we use the hard X-rays (2-8 kev) to search for
the possible presence of an obscured AGN. In every case, a hard X-ray
source is detected in the nuclear region. 
If we divide the sample 
according to the optical/IR spectroscopic classification (starburst
{\it vs.} AGN), we find that the 5 ``starburst'' ULIRGs have hard X-ray
luminosities about an order-of-magnitude smaller than the 3 ``AGN''
ULIRGs. NGC 6240 has an
anomalously high hard X-ray luminosity compared to the ``starburst'' ULIRGs.
The Fe K$\alpha$ line is convincingly detected in only two ULIRGs.
The weakness of
the Fe-K emission in these ULIRGs generally suggests that the hard X-ray
spectrum is not dominated by reflection from high $N_H$ neutral 
material. 
The hard X-ray continuum flux ranges from a few $\times 10^{-3}$ to
a few $\times 10^{-5}$ of the far-IR flux, similar to values in pure
starbursts, and several orders-of-magnitude smaller than in Compton-thin AGN.
The upper limits on the ratio of the Fe K$\alpha$ to far-IR flux are  
below the values measured
in Compton-thick type 2 Seyfert
galaxies.  
While very large column densities of molecular gas are
observed in the nuclei of these galaxies, we find no evidence that
the observed X-ray sources are obscured by Compton-thick material.
Thus, our new hard
X-ray data do not provide direct evidence that powerful ``buried
quasars'' dominate the overall energetics of most ultraluminous infrared
galaxies.
\end{abstract}


\section{Introduction}
Ultraluminous IR galaxies (ULIRGs) are defined to be galaxies whose
bolometric luminosity exceeds $10^{12} L_{\odot}$ and whose emission
is dominated by the mid- and far-infrared bands \citep{sa96}.
\footnote{We assume $H_0 = 70\rm \ km \ s^{-1} \
Mpc^{-1}$.} They are thought to be powered by either an AGN 
%(Sanders 1999)
\citep{sa99}
, a very powerful nuclear starburst \citep{jo99}, or both. It is
possible that these galaxies are an evolutionary stage of normal
galaxies in which a large amount of dense gas that has been driven to
the nucleus is both feeding a nuclear black hole and fueling a very
high star-formation rate %(e.g. Scoville \& Norman 1988)
\citep{ns88}.

ULIRGs generically have highly disturbed morphologies suggestive of
severe tidal effects, and often have double nuclei suggestive of the
late stages of a galaxy merger \citep{cl96}. If elliptical galaxies are
the by-product of similar mergers, then ULIRGs should have been very
common in the early universe \citep{ge01}. This is consistent with the
properties of the sub-mm source population, and the corresponding
estimates for the star-formation history of the universe \citep{bl99}.

ULIRGs offer ideal local laboratories for studying the processes
occurring in the high-redshift sub-mm sources. As such, decoupling the
effects of an AGN and star-formation is important for assessing the
role of ULIRG-like objects in building present-day elliptical
galaxies, in accounting for the total amount of energy supplied by AGN
over cosmic time, and in establishing the strong correlation between
the mass of a supermassive black hole and the properties of its "host"
galaxy spheroid %(e.g. Ferrarese \& Merritt 2000; Tremaine et al. 2002).
\citep{fe00, tr02}.

The nature of the central power-source in ULIRGs remains uncertain,
owing to its location behind large column densities of obscuring
material. This highlights the importance of observations in the hard
X-ray, far-IR, and radio domain, which most effectively pierce the
obscuring material. However, such observations to date have not
provided decisive evidence to break the AGN/starburst deadlock
(because they have lacked either the appropriate spatial resolution or
have not provided unambiguous spectroscopic discriminants between AGN
and starbursts).

The optical and near/mid-IR regimes contain many potentially important
spectral diagnostics. Such observations of ULIRGs variously classify
them as Seyferts, LINERs or starbursts \citep{ge98, ke01, im00, la00,
v99a,v99b, kim98a, v95}.  If the ULIRGs optically classified as LINERs
are considered to be starburst-driven, then there is good overall
agreement between the optical and IR classifications \citep{lu99,
ta99}
, and in general nearby LINERs exhibiting only narrow H$\alpha$ lines tend to have X-ray properties more consistent with a starburst rather than an AGN interpretation \citep{te00, er02}. However, X-ray observations of the
infrared-luminous merger NGC 6240 (which appears to be starburst-dominated
LINER from an optical/IR perspective) reveal a powerful AGN that may
dominate its energetics 
%(e.g. Matt et al. 2000).
%\citep{ma00}.
\citep{vi99}.

Observations of ULIRGs with the Chandra X-ray Observatory promise to
be enlightening for this issue. Chandra has large effective area, high
spatial resolution (FWHM $\sim$ 1\arcsec), and is capable of moderate
spectral resolution imaging spectroscopy. All of these may be
necessary to determine the physical origin of the hard X-ray emission
previously detected from ULIRGs 
%(Risaliti et al. 2000). 
\citep{ri00}.
Recently,
%Gallager et al. (2002)
\citet{ga02},
%Xia et al. (2002)
\citet{xi02}, and \citet{cl02}
have presented detailed
analysis of Chandra observations of the ULIRGs Mkn 231, Mkn 273, and Arp 220
respectively \citep[see also][]{mc03}. Our approach here is to examine a sample of ULIRGs that
is large enough to document the properties of the class as-a-whole.

The eight ULIRGs listed in Table 1 comprise both a
flux-limited sample with 
$F_{60\mu} > 10$ Jy, and a volume-limited sample out to cz = 14000 km s$^{-1}$
(D $<$ 200 Mpc), and  NGC 6240 is included since it is an archetype of 
IR-luminous galaxies shown by hard X-ray data to
harbor a powerful AGN \citep{iw98}.  
Note that the galaxies are all at a redshift of $\sim 
0.04$ with the exceptions of Arp 220 and NGC 6240 ($z \sim 0.02$).
The joint analysis of XMM-Newton and Chandra data for two of these
galaxies (IRAS 05189-2524 and UGC 05101) will be presented in future
work. An analysis of the  
spatially-resolved soft X-ray emission and a comparison to models of 
galactic winds will also be presented elsewhere.

The organization of the present paper is as follows. In section 2 we
describe the data reduction and analysis techniques. In section 3 we discuss
the results of spatial and spectral fitting of the data. In section 4 we
discuss the implications of these results.

%\clearpage
\begin{deluxetable*}{lllllll}
\tabletypesize{\scriptsize}
\tablecaption{Basic data for the ULIRG sample.}
\tablehead{
\colhead{Galaxy} & \colhead{Position} &
\colhead{z} & 
\colhead{Scale}&
%\colhead{Satellite} & 
\colhead{Date} & \colhead{Exposure}
& \colhead{Galactic $N_H$}\\
%\colhead{$F_{25\mu m}$} & \colhead{$F_{60\mu  m}$} & 
%\colhead{$F_{100 \mu m}$} \\ 
& \colhead{(J2000)} &  & \colhead{(kpc/\arcsec)} & &  \colhead{(ks)} &
\colhead{($10^{20} \rm \ cm^{-2}$)}}
\startdata
IRAS05189-2524 & 05 21 01.5  -25 21 45 & 0.0426 & 0.82 & 
%0.89 & XMM-Newton & 03/17/2001 & 8.1 \\
%& 2 \\ 
%& & & & 
%Chandra & 
10/30/2001 & 6.6 & 4 \\% IRAS 3.5 & 14.0 & 12.5
& & & & 
%Chandra & 
01/30/2002 & 14.8 \\
UGC 05101 & 09 35 51.6  +61 21 11 & 0.0394 & 
0.76 & 
%Chandra & 
05/28/2001 & 49.3 & 3 \\
% IRAS  1.0 & 11.5 & 20.2
%& & & & XMM-Newton & 11/12/2001 & 26.7 \\
Mkn 231 & 12 56 14.2  +56 52 25 & 0.0422 & 
0.82 & 
%Chandra & 
10/19/2000 & 34.5 & 1\\
% IRAS  8.7 & 32.0 & 30.3 \\
Mkn 273 & 13 44 42.1  +55 53 13 & 0.0378 & 
0.74 & 
%Chandra & 
04/19/2000 & 44.2 & 1\\
% IRAS  2.3 & 21.7 & 21.4 \\
Arp 220 & 15 34 57.1  +23 30 11 & 0.0181 & 0.36 & 
%Chandra & 
06/24/2000 & 56.4 & 4\\
% IRAS 7.9 & 103.8 & 112.4 \\
NGC 6240 & 16 52 58.9 +02 24 03 & 0.0245 & 0.49 & 07/29/2001 &
36.6 & 6\\
IRAS17208-0014 & 17 23 21.9  -00 17 00 & 0.0428 &
0.83 & 
%Chandra & 
10/24/2001 & 48.5 & 10\\
% IRAS 1.7 & 31.1 & 34.9 \\ 
IRAS20551-4250 & 20 58 26.9 -42 39 0 & 0.0428 & 
0.83 & 
%Chandra & 
10/31/2001 & 40.4 & 4\\
% IRAS 1.9 & 12.8 & 10.0 \\
IRAS23128-5919 & 23 15 47.0  -59 03 17 & 0.0446 & 
0.86 & 
%Chandra & 
09/30/2001 & 33.8 & 3\\
% IRAS 1.6 & 10.8 & 11.0 \\
\enddata
\tablecomments{Positions and redshifts
%and IRAS fluxes 
were obtained from NED.  Galactic $N_H$ values are determined from
\citet{di90} using the
LHEASOFT tool ``nh'' (http://heasarc.gsfc.nasa.gov)\\
}
\end{deluxetable*}
%\clearpage
\section{Data Reduction}
The observation dates and net exposure times for our targets are listed
in Table 1.  The data for Arp 220, Mkn 231, Mkn 273 and NGC 6240 were
downloaded from the Chandra archive.  The data
reduction was performed using CIAO 2.2 and the data were reprocessed
using CALDB 2.9.  XAssist \citep[see][]{pt02},
which is a software package that assists in data reprocessing,
initial source selection and analysis, was also used in the
analysis of these data. 
Briefly,
XAssist performs the basic data reduction steps recommended by the CXC
``threads'' (note that the 0.5 pixel position randomization is also removed,
which is an optional step).
Sources are detected using wavdetect, the background light curve is examined
and time of background flaring are removed, and the image of each source is
fitted with an elliptical gaussian model in order to determine the spatial
extent (see section 3).  
The data were analyzed in full
(0.3-8.0 keV), soft (0.5-2.0 keV) and hard (2.0-8.0 keV) bandpasses.
For comparison with previous X-ray studies a region 
was selected (manually) that encompassed the entire X-ray extent of the galaxy
(determined from the full-band images) and spectra were extracted for
these ``global'' regions.  The spectra were binned to 20 counts/channel to allow
use of the $\chi^2$ statistic.

Figure 1 shows adaptively-smoothed
soft and hard band 
images, with the global and nuclear regions marked. We show the soft
band images here for completeness, but defer their analysis. Note that
the nuclear regions were chosen to maximize the signal-to-noise 
of the nuclear spectrum
while the
emphasis for the global regions was complete coverage of the X-ray
emission of the galaxies (in part for comparison with previous X-ray
studies).  We checked for counterparts to all
X-ray sources within the central 8' of the observation (where the PSF
size is $< 1\arcsec$) using HEASARC databases containing stars (USNO
and GSC2.2), QSOs, 2Mass sources and the FIRST survey.  In
general, only a few counterparts in each field were found. These show
that the astrometric solutions are good to $\sim 
0.5-1\arcsec$, but could not be used to apply reliable astrometric
corrections.  

\section{Spatial Analysis}

\subsection{Nuclear Source Sizes}
The extents derived from the spatial fitting procedure are listed in
Table 2.  The fitting was performed using the C fit statistic
\citep{ca79} and the errors correspond to $\Delta C = 4.605$ (90\%
confidence interval for 2 interesting parameters). In all but three
cases, the nuclear 2.0-8.0  
keV source is consistent with being unresolved (FWHM $\lesssim 1 \arcsec$).
The hard X-ray emission in the nuclei of both IRAS 17208-0014 and Arp 220
are resolved and asymmetric, with a major axis FWHM of 2.6 $\arcsec$
(0.9 kpc) in Arp 220 and 4.7 $\arcsec$ (3.9 kpc) in IRAS 17208-0014.
Point sources are present in the FOV with Gaussian fit FWHM
parameters of $\sim 1$\arcsec,
indicating that the elongation in IRAS 17208-0014 is not due to errors
in the aspect solution. 
In NGC 6240, the hard X-ray nuclear emission is resolved into two
sources coincident with the radio nuclei, and
diffuse emission, consistent with the full-band HRC
results reported in \citet{li02b} and the ACIS results reported in
\citet{ko03}.  However for consistency with the
rest of the sample we model the hard X-ray flux with a single Gaussian
to determine the extent of the total (i.e., combined) nuclear flux,
noting that the
individual nuclei would not have been resolved at redshifts 
$\ga 0.04$ (see also below).

%\clearpage
\begin{deluxetable*}{llllllll}
\tabletypesize{\scriptsize}
\tablecaption{Nuclear Spatial Properties}
\tablehead{
\colhead{Galaxy} & \multicolumn{3}{c}{Soft Emission} &
\multicolumn{3}{c}{Hard Emission} \\
\colhead{} & 
\colhead{$\sigma_x$ (\arcsec)} & \colhead{$\sigma_y$ (\arcsec)} &
\colhead{$\theta$} & 
\colhead{$\sigma_x$ (\arcsec)} & \colhead{$\sigma_y$ (\arcsec)} &
\colhead{$\theta$}}
\startdata
%Arp 220 & 0.97 (0.96-0.98) & 1.09 (1.07-1.11) & 167 (161-178) & 0.97
%(0.83-1.12) & 0.65 (0.55-0.78) & 10 (-5 - -27) \\
% ---- Arp 220 soft x-ray values updated 9/11/02
Arp 220 & 1.05 (1.03-1.06) & 0.97 (0.95-0.98) & 64 (59-67) & 
% 4/11/03 added 180 to rot.
%0.99 (0.86-1.14) & 0.68 (0.57-0.82) & 10 (-5 - 30) \\
0.99 (0.86-1.14) & 0.68 (0.57-0.82) & 190 (175 - 210) \\
%IRAS05189-2524 & 0.46 (0.41-0.50) & =$\sigma_x$ & \nodata &
%0.44 (0.41-0.47) & =$\sigma_x$ & \nodata \\
IRAS05189-2524 & 0.46 (0.42-0.49) & =$\sigma_x$ & \nodata &
0.44 (0.41-0.47) & =$\sigma_x$ & \nodata  \\
IRAS05189-2524\tablenotemark{*} & 0.50 (0.46-0.53) & =$\sigma_x$ & \nodata &
0.46 (0.44-0.48) & =$\sigma_x$ & \nodata \\
IRAS17208-0014 & 2.02 (1.62-2.53) & 1.45 (1.20-1.76) & 57 (34-81) &
1.05 (0.89-1.26) & 0.58 (0.48-0.74) & 66 (54-81) \\
%IRAS17208-0014 & 1.5 ($>$1.5) & =$\sigma_x$ & \nodata &
%0.57 (0.45-0.72) & 1.06 (0.89-1.27) & 28 (17-42) \\
%IRAS20551-4250 & 0.93 (0.80-1.25) & 1.40 ($>1.21$) & 51 (30-163) &
%0.64 (0.53-0.84) & =$\sigma_x$ & \nodata\\
IRAS20551-4250 & 1.53 (1.15-1.86) & 1.06 (0.76-1.33) & 150 (127-170) &
0.65 (0.53-0.84) &  =$\sigma_x$ & \nodata\\
%IRAS23128-5919 & 1.21 ($>0.81$) & 0.67 (0.51-0.91) & -38 (-55 -
%-17) & 
%0.43 (0.38-0.48) & =$\sigma_x$ & \nodata \\
IRAS23128-5919 & 1.35 (0.94-1.70) & 0.77 (0.55-1.06) & 146 (131-168) &
0.43 (0.38-0.48) &  =$\sigma_x$ & \nodata\\
%Mkn 231 & 0.43 (0.41-0.45)& =$\sigma_x$ & \nodata &
%0.42 (0.39-0.44) & =$\sigma_x$ & \nodata \\
% fixed 4/11/03
%Mkn 231 & 0.43 (0.38-0.48 )& =$\sigma_x$ & \nodata &
Mkn 231 & 0.43 (0.40-46)& =$\sigma_x$ & \nodata &
0.42 (0.40-0.44) &  =$\sigma_x$ & \nodata \\
%Mkn 273 & 0.39 (0.37-0.41) & =$\sigma_x$ & \nodata &
%0.59 (0.55-0.64) & 0.47 (0.43-0.50) & -40 (-52 - -28) \\
Mkn 273 & 1.17 (1.06-1.29) & 0.61 (0.53-0.69) & 45 (38-52) &
% fixed 4/11/03
%0.59 (0.55-0.64) & 0.47 (0.53-0.50) & 141 (128-152) \\
0.59 (0.55-0.64) & 0.47 (0.43-0.50) & 141 (128-152) \\
%UGC 05101 & 0.97 (0.83-1.10) & =$\sigma_x$ & \nodata &
%0.53 (0.46-0.61) & =$\sigma_x$ & \nodata\\
NGC 6240 & 1.85 (1.72-1.99) & 2.79 (2.58-3.04) & 158 (150-166) & 0.80
(0.72-0.88) & 1.19 (1.12-1.26) & 14.5 (8.9-21.4) \\ 
UGC 05101 & 0.97 (0.83-1.10) & =$\sigma_x$ & \nodata &
0.55 (0.46-0.63) & =$\sigma_x$ & \nodata\\
\enddata
\tablecomments{Fit results based on fitting an elliptical gaussian model
where surface brightness = 
$Ne^{-[(\frac{x}{2\sigma_x})^2 + (\frac{y}{2\sigma_y})^2]}$, and x, y
are rotated from the right 
ascension axis by $\theta$ (given in
degrees). Note that these fits were performed only {\it to the nuclear
  source} in images extracted 
from the central 5\arcsec of these galaxies, and as shown in Figure 1 often
emission is present on larger spatial scales.}
\tablenotetext{*}{2nd observation of IRAS 05189-2524.}
\end{deluxetable*}
\begin{figure*}[htbn]
\epsscale{1.}
\plotone{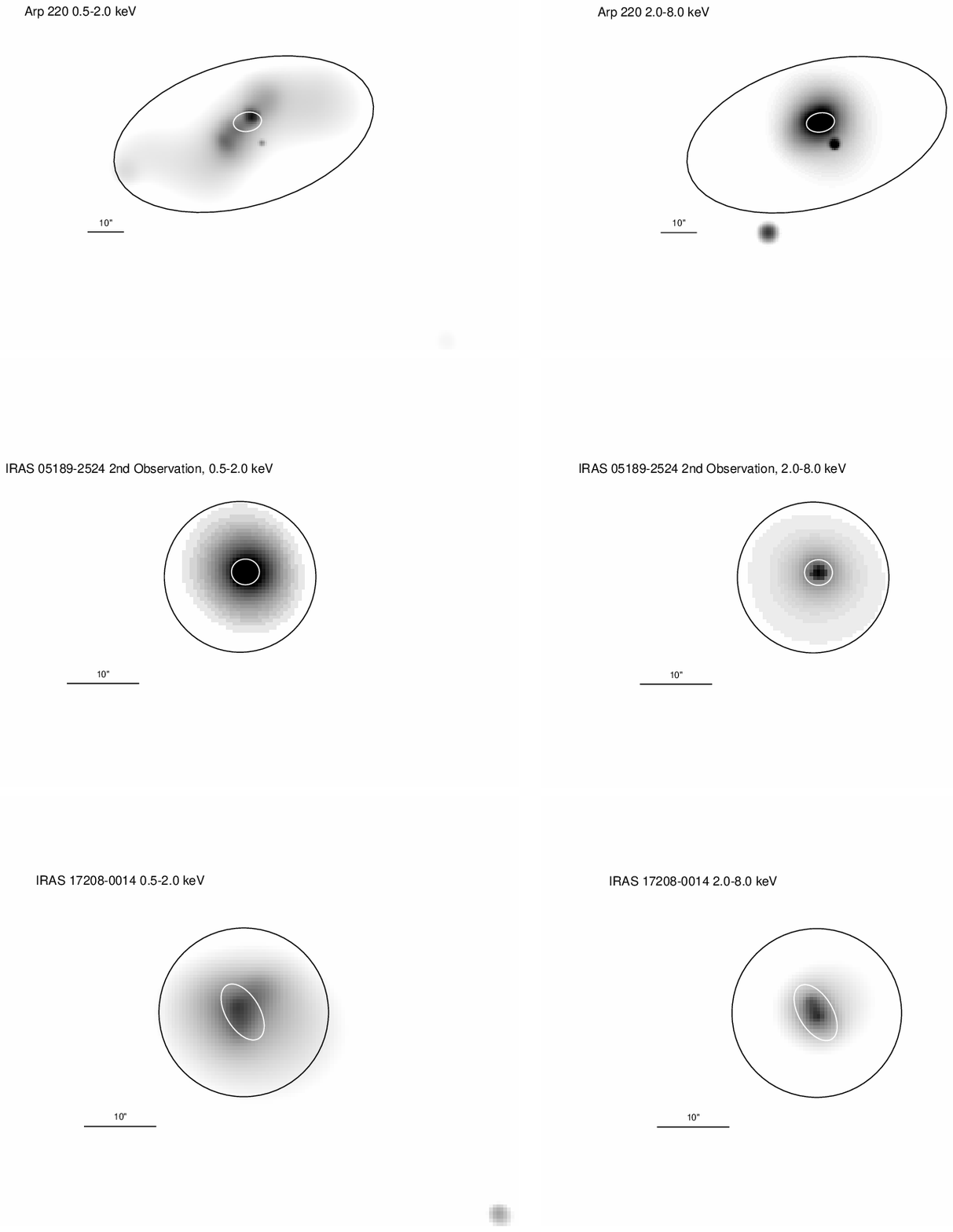}
\caption{} 
\end{figure*}
\setcounter{figure}{0}
\begin{figure*}[htbn]
\plotone{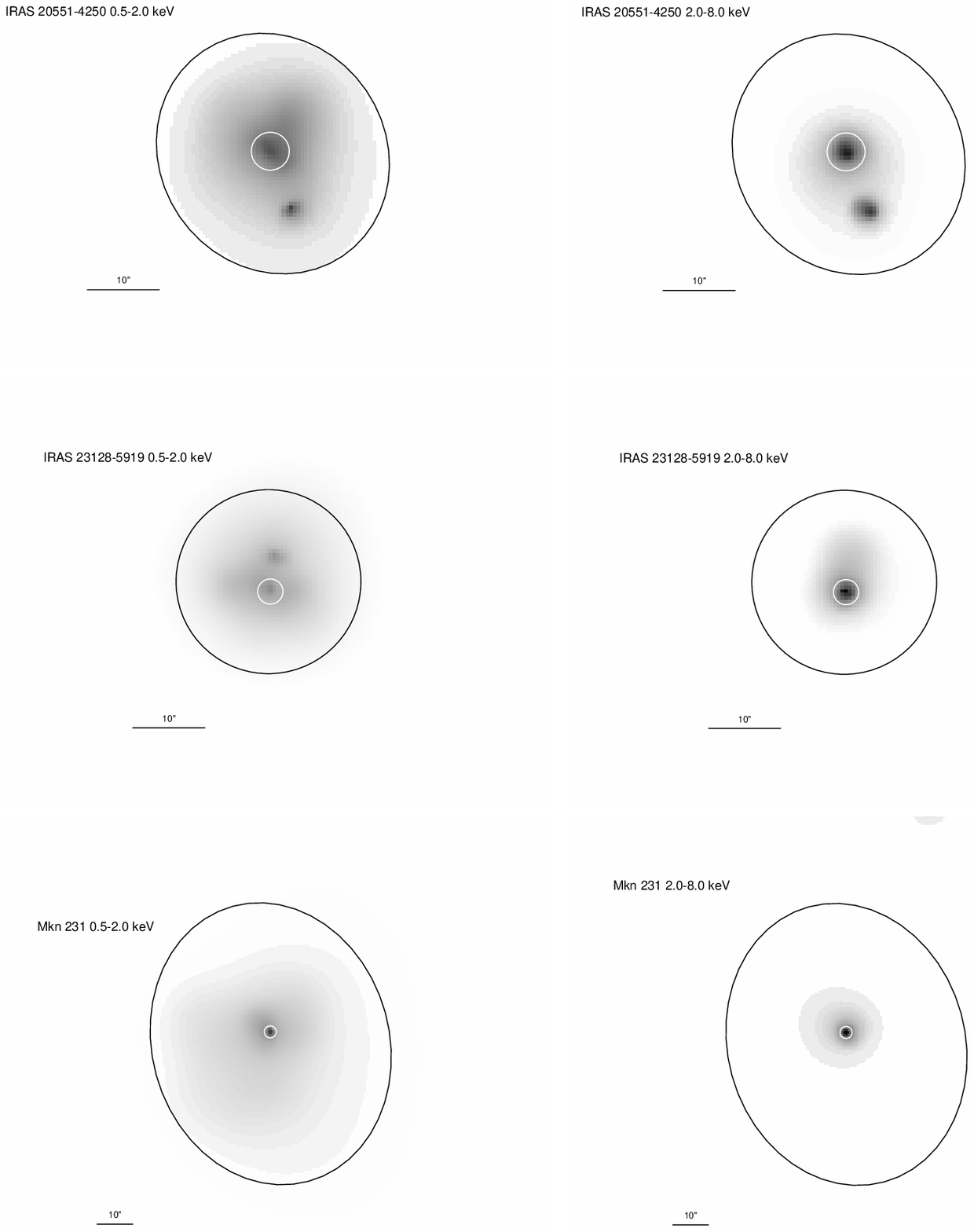}
\caption{(cont.)}
\end{figure*}
\setcounter{figure}{0}
\begin{figure*}[htbn]
\plotone{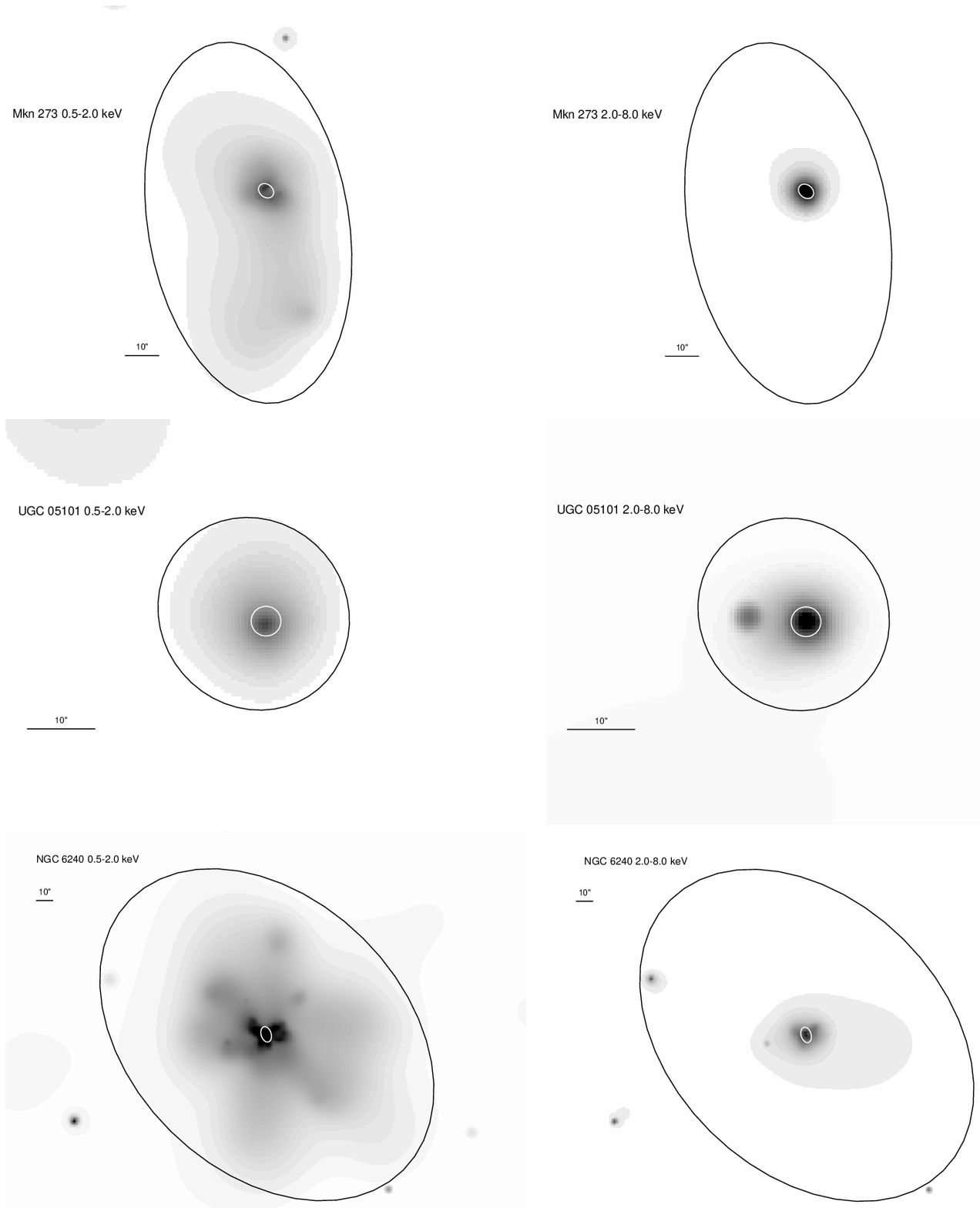}
\caption{(cont.) 0.5-2.0 keV and 2.0-8.0 keV images of the ULIRG sample.  The
outer and inner ellipses show the global and nuclear source region,
respectively.  The images were adaptively smoothed to a significance
level of 2.8$\sigma$ and scaled logarithmically.  N is up and E is to
the left in these images.
} 
\end{figure*}
%\clearpage

%\clearpage
In order to further investigate the presence of unresolved hard X-ray
emission the 2-8 keV nuclear emission was fit with a Gaussian model
convolved with a PSF image generated using the CIAO tool mkpsf (using
a photon energy of 3 keV, where the 2-8 keV photon spectrum generally
peaks for these sources).  For the galaxies with a nuclear point
source, the main impact of including the PSF was to reduce the source
extent parameter $\sigma$ from $\sim 0.5$\arcsec to $\sim
0.25$\arcsec: that is, the model PSF convolved with a Gaussian with
$\sigma$=0.25\arcsec (FWHM = 0.6 arcsec) provides a good fit to the
nuclear point sources in these ULIRGs. We regard this as an upper
limit to the true source size, since it may arise from errors in the
aspect solution.
%This implies that the combination of the difference between the
%true and modeled PSF and any error in the aspect solution is
%effectively modeled with a Gaussian with $\sigma \sim 0.25\arcsec$. 
%This suggests that the Chandra PSF and any systematic error in this
%fitting process (e.g., residual aspect errors and any error in the PSF
%model) are each contributing on the order of
%$\sim 0.25\arcsec$ to the observed PSF

As stated above, NGC 6240 is resolved into two faint sources in
addition to diffuse flux.  Modeling this flux as a combination of two
point sources in addition to two elliptical Gaussians (to account for
diffuse flux associate with each nuclei) we find that the northern and
southern nuclei contribute 17\% and 28\% of the 2-8 keV counts from
the central 9$\arcsec$ ($\sim 0.5$ kpc) of NGC 6240.   
In the cases of Arp 220 and 
IRAS 17208-0014, the diffuse emission was modeled with elliptical
Gaussians and the PSF $\times$ Gaussian model with $\sigma = 0.25\arcsec$
was placed at the centroid of the diffuse flux.  This resulted in
upper limits (for $\Delta C = 4.605$) of $1.2 \times 10^{-3}$ and $4.3
\times 10^{-4}$ counts s$^{-1}$ for the 2-8 keV count rate of a point
source, respectively, or 2-10 keV fluxes of $4.5 \times 10^{-14}$ \ergcms and
$8.3 \times 10^{-15}$ \ergcms\ ($L_{2-10 \rm \ keV} \sim 4 \times
10^{40}$ \ergs\ in both cases).  Finally, we note that in Arp 220 there
is a $\sim 1.7\arcsec$ (0.6 kpc) offset between the hard and soft nuclear
positions, with the hard nucleus being coincident with the western 6 cm
nucleus discussed in \citet{no88} and the soft nucleus being offset
to the NW of the hard nucleus \citep{cl02}. The soft nucleus is not
coincident 
with either radio nucleus and is most likely due to hot ISM observed
through a region with relatively low extinction. Note that the optical
nucleus of Arp 220 is similarly offset by $\sim 2''$ to the NW of the
western radio nucleus, which has likewise been interpreted as the result of a
``hole'' in the extinction toward the nucleus \citep{ar01}.  
%More detailed
%spatial analysis will be given in Paper III (Heckman et al. 2002, in
%preparation).

\subsection{Extra-nuclear Point Sources}
Table \ref{ixotab} lists the positions and luminosities of unresolved
sources that were detected within the X-ray extent of the galaxies.
The luminosities were computed assuming a power-law model with a
photon index of 1.8.
If these sources are associated with the corresponding host galaxy,
then they have luminosities well in excess of the Eddington luminosity
of a solar-mass black hole or neutron star X-ray binary, and hence are
``ultraluminous X-ray objects'' (ULXs; also known as
intermediate-luminosity X-ray objects or IXOs). Note that the ULX
in UGC 05101 has a very hard spectrum (i.e., it is not present in the
soft-band image in Figure 1) which may be indicative of a highly-absorbed
spectrum with $N_H > 10^{22-23}\ \rm cm^{-2}$.  Since this source is
not in the nucleus of UGC 05101, this would imply a large amount of
internal absorption for this source.  \citet{ga02}
also discuss an ULX in Mkn 231 which lies below our detection
threshold of $\sim 10$ counts.  Finally, the ULX in IRAS 20511-4250 has an
estimated hard X-ray luminosity that is $\sim 20\%$ of the luminosity
of the nuclear source in this galaxy.  However, this luminosity is
highly uncertain given the low count rate of the source.  The 2-8 keV
count rate of the ULX is $\sim 25\%$ of global 2-8 keV count rate. 

\clearpage 
\begin{deluxetable*}{lllllll}
\tablecaption{Extra-nuclear Point Sources\label{ixotab}}
\tabletypesize{\scriptsize}
\tablehead{
\colhead{Galaxy} & \colhead{Position} & \colhead{Offset} & 
\colhead{Counts} & \colhead{$F_{2-10 \ \rm keV}$} & 
\colhead{$L_{2-10 \ \rm keV}$} & \colhead{Notes} \\
\colhead{} & \colhead{} & \colhead{(\arcsec/kpc)} & \colhead{} &
\colhead{$10^{-15}$ \ergcms} & 
\colhead{$10^{40}$ \ergs}
}
\startdata
Arp 220 & 15 34 57.0 23 30 5.6 & 7.3/2.8& 29.2 & 1.8 & 0.13\\
IRAS 20551-4250 & 20 58 26.5 -42 39 8.6 & 8.4/7.5 & 87.6 & 6.9 & 2.8 \\
IRAS 23128-5919 & 23 15 46.7 -59 3 11.0 & 4.7/4.0& 75.5 & 6.4 & 2.9 &
Northern nucleus\\
UGC 05101 & 9 35 52.7 61 21 12.3 & 8.3/6.8 & 17.5 & 1.2 & 0.42 \\
\enddata
\end{deluxetable*}
%\clearpage

\subsection{Large-Scale Emission}

In addition to the nuclear source and the extranuclear point sources,
extended hard X-ray emission is often detected beyond
the nuclear region on scales of $\sim$ 10 kpc. The fraction of total
hard X-ray flux in this spatially-extended component is $\leq$ 10\% in
four cases (Arp 220, IRAS 05189-2524, Mkn 273, and Mkn 231), but
ranges from 30 to 50\% in the other cases (Figure 1).  Note that
in the case of IRAS 20551-4250, the extra-nuclear unresolved source may
account for up to $\sim 50\%$ of the extra-nuclear hard X-ray flux.

\section{Spectral Analysis}
% --- 9/13
%\subsection{Nuclear Spectra}
As stated above, in every galaxy a nuclear core was detected in the
2-8 keV data.  We
proceeded initially by using XSPEC to fit these spectra with a simple absorbed
power-law model.  The results of this
procedure are given in Table 4.  Errors (corresponding to
$\Delta\chi^2=4.605$) are only derived for fits in which $\chi^2$/dof
$< 1.5$.  Figure 2 shows the data and model for these fits.
These fits result in $\chi^2$/dof $<$
1.5 only in the cases of IRAS 17208-0014 and Mkn 231.

%\clearpage
\begin{figure*}[htbn]
\plotone{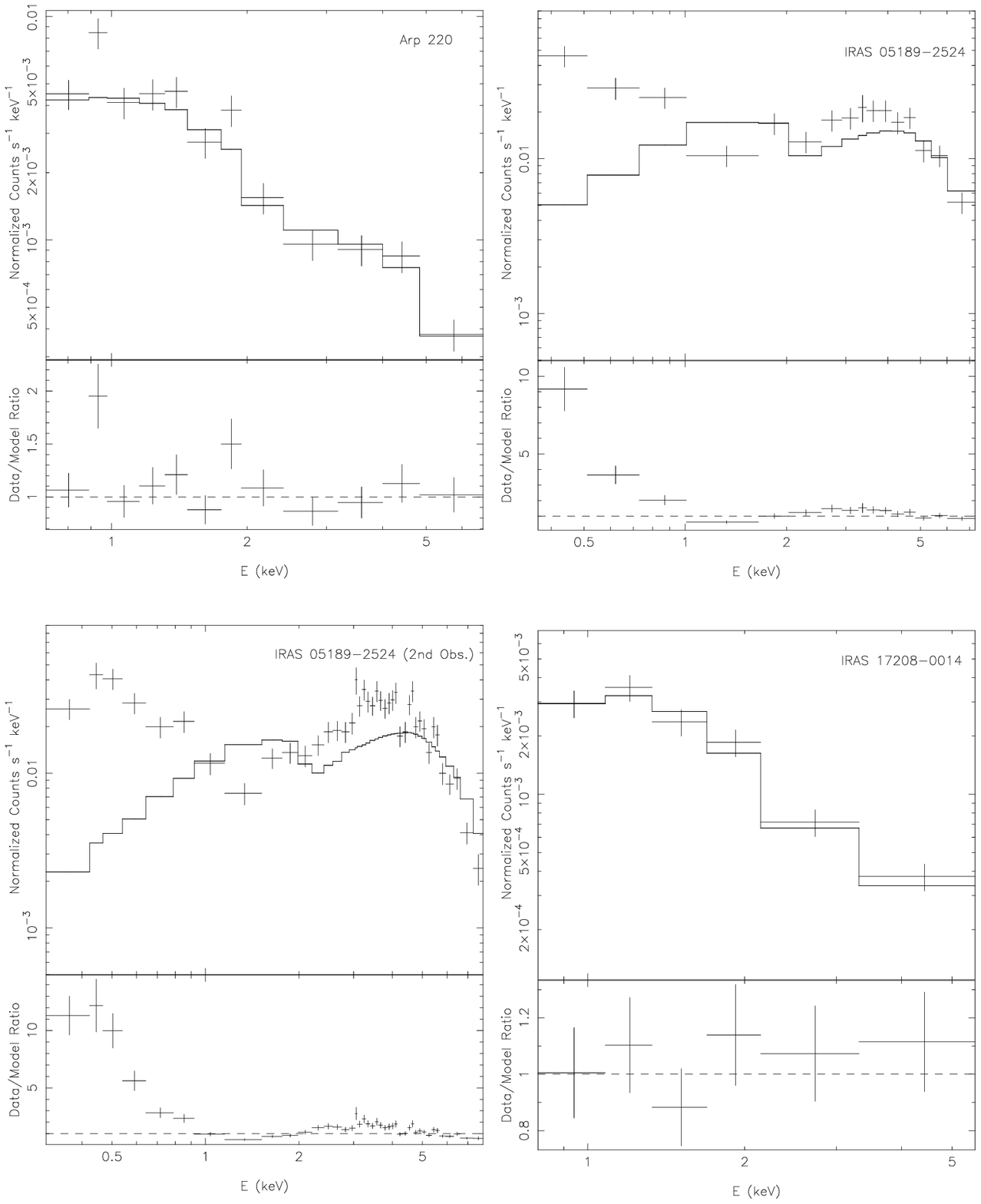}
\caption{}
\end{figure*}
\setcounter{figure}{1}
\begin{figure*}[htbn]
\plotone{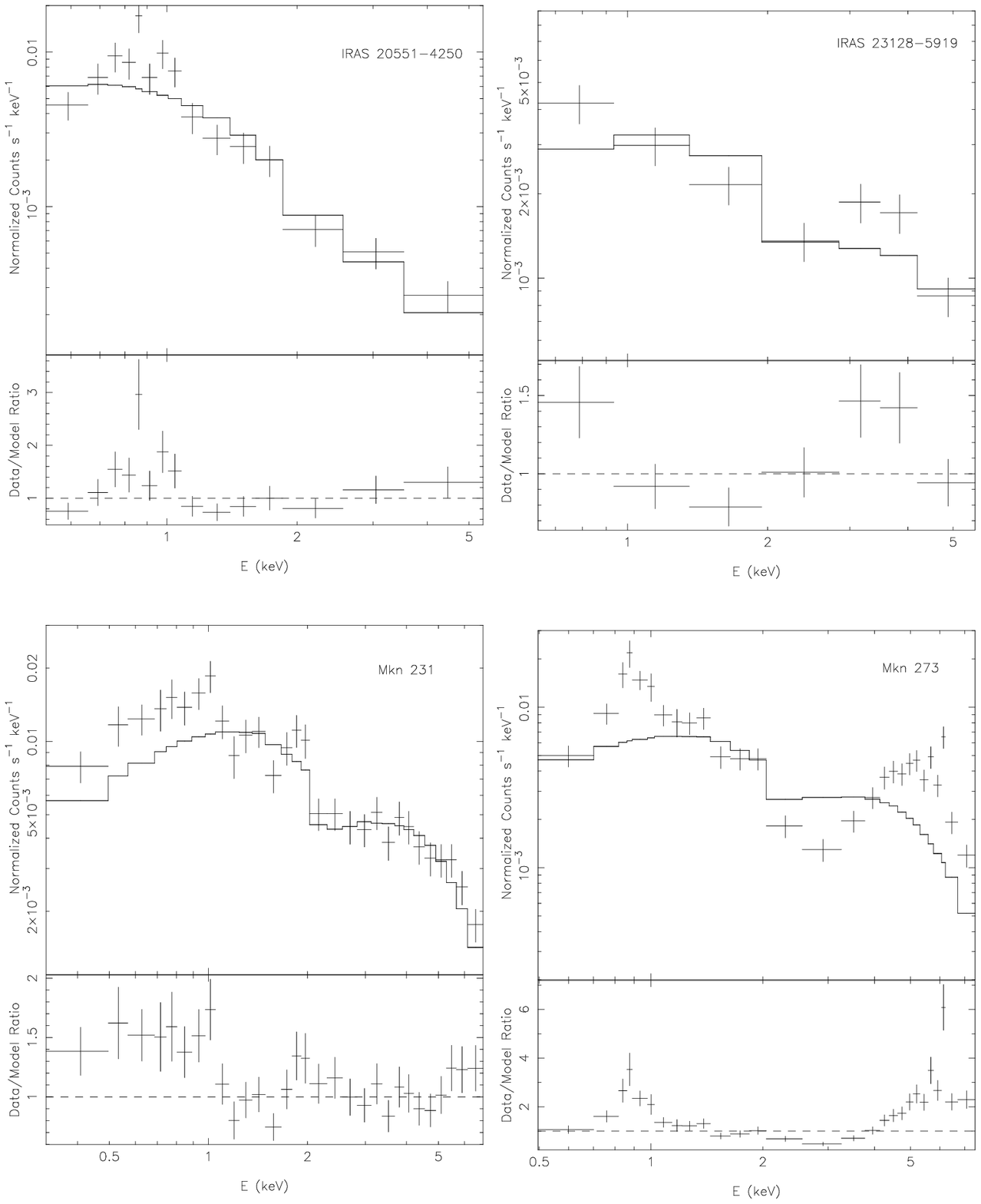}
\caption{(cont.)}
\end{figure*}
\setcounter{figure}{1}
\begin{figure*}[htbn]
\plottwo{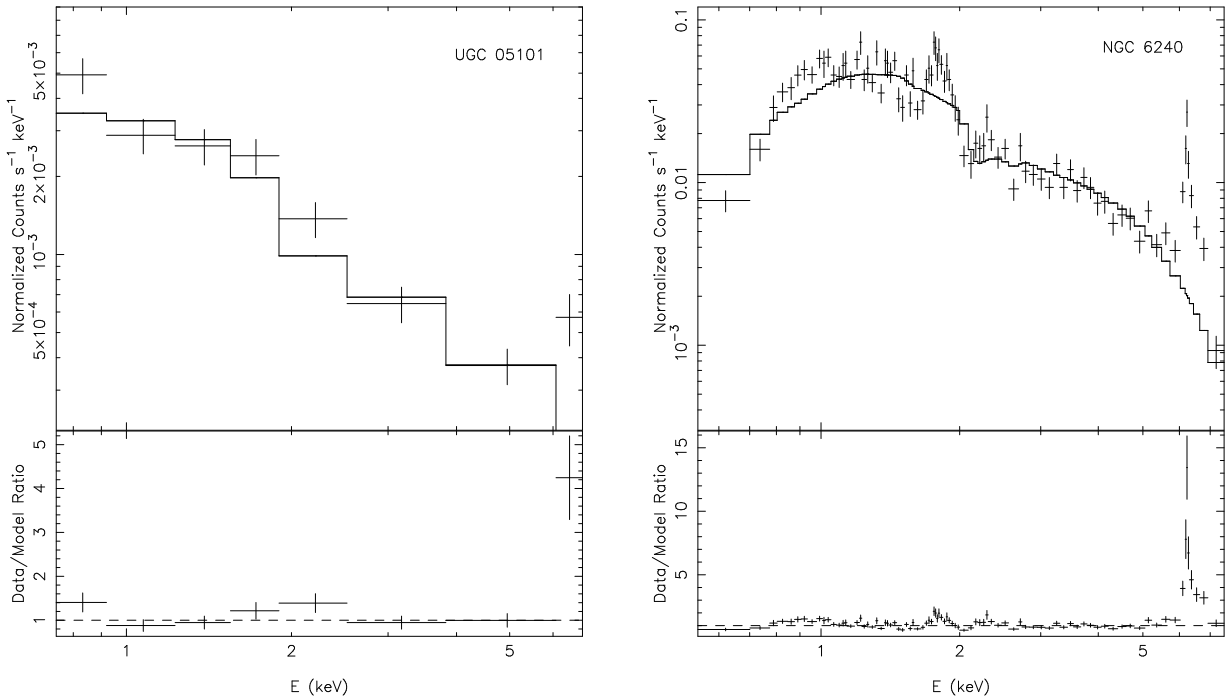}{f2j.eps}
\caption{Power-law fits to the nuclear spectra of the ULIRG galaxies.
The top panel shows the data and model and the data/model ratio is
shown in the bottom panel.}
\end{figure*}
%\clearpage

\begin{deluxetable}{llll}
\tablecaption{Power-law Fits to Nuclear Spectra}
\tabletypesize{\scriptsize}
\tablehead{
\colhead{Galaxy} & \colhead{$N_H \ (10^{22} \ \rm cm^{-2})$} &
\colhead{$\Gamma$} & \colhead{$\chi^2$/dof}
}
\startdata
% All values revised 4/10
Arp 220 & 0.0 & 0.9 & 39.5/21\\
IRAS 05189-2524 & 0.0 & -0.4 & 127/28 \\
IRAS 05189-2524\tablenotemark{*} & 0.0 & -0.7 & 430.2/78 \\
IRAS 17208-0014 & 0.3 ($<$ 0.6) & 1.7 (1.0-2.3) & 10.5/9 \\
IRAS 20551-4250 & 0.1 & 2.0 & 30.9/12 \\
IRAS 23128-5919 & 0.0 & 0.4 & 20.4/11 \\
Mkn 231 & 0.01 ($<$0.02) & 0.44 (0.33-0.53) & 57.0/53 \\
Mkn 273 & 0.0 & 0.3 & 302/49 \\
UGC 05101 & 0.0 & 1.0 & 35.5/12 \\
NGC 6240 & 0.39 & 1.4 & 431/136 \\ % n6240_nucl_cl_pl.log
\enddata
\tablenotetext{*}{2nd observation}
\end{deluxetable}
%\clearpage

We next tried to fit each spectrum with a ``mekal'' plasma \citep{li95}
plus power-law model.  In these fits, absorption is
applied to the entire spectrum with a hard lower-limit on the column
density set to be 50\% of the Galactic column, and a second absorber
at the redshift of the ULIRG is applied only to the power-law
component.  Based on X-ray spectroscopy of starburst galaxies, the
temperature of the plasma was limited to the range 0.3-2.0 keV. If the
temperature was not constrained by the data then it was held fixed at
0.7 keV.  Likewise, the abundance was limited to the range of $Z=0-5
Z_{\odot}$, and fixed at $1.0 Z_{\odot}$ if not constrained.  The
results are given in Table 5 and plotted in Figure 3.  Here all fits have $\chi^2$/dof $<1.5$
except in the cases of Mkn 273, UGC 05101 and NGC 6240 (due to Fe-K
emission, see below).  Note that this more complex model is statistically
preferred to the simple power-law model for Mkn 231.
%XXX Give stat. prob. of observing $chi^2$ given dof
%rather than going on just $chi^2$/dof $< 1.5$ since here stat. errors
%dominate over systematic errorsXXX.
% and residuals are 
%also evident around 6.4 keV in UGC 05101 although they are
%not highly significant (see below).  
Significant residuals are evident in the case of Arp 220 around
0.8 and 1.8 keV, possibly the result of non-solar O and Si abundances
or additional thermal components at different temperatures being
required. A degeneracy between these scenarios is often observed in
nearby starburst galaxies \citep[e.g.,][]{we00}.

Chandra observations of bright point sources may suffer from pile-up
\citep{da01}. 
This is only a concern for IRAS 05189-2524, the brightest ULIRG in
our sample and the only one that is completely unresolved at both
soft and hard X-ray energies \citep[see also the pile-up discussion
concerning Mkn 231 in][]{ga02}. We assessed the amount of pile-up by
modeling the 2nd observation, 
which had 
the higher count rate, using MARX.  This analysis indicated that $\sim 20\%$
of the Chandra counts are piled-up in that observation.  We checked the impact
by fitting the two spectra from the two observations of IRAS 05189-2524
with the
plasma plus power-law model, additionally including the ``pile-up'' convolution
model in XSPEC \citep[based on][]{da01}.  We allowed the overall normalization
and the pile-up grade migration parameter to vary independently.  This
resulted in a power-law slope of $\Gamma$ = 1.1, consistent 
with the individual fits, and showed that we are underestimating the 2-10 keV
flux of IRAS 05189-2524 by $\sim 15\%$.  Therefore pile-up does not affect
our conclusions.

%\clearpage
\begin{figure*}[htbn]
\plotone{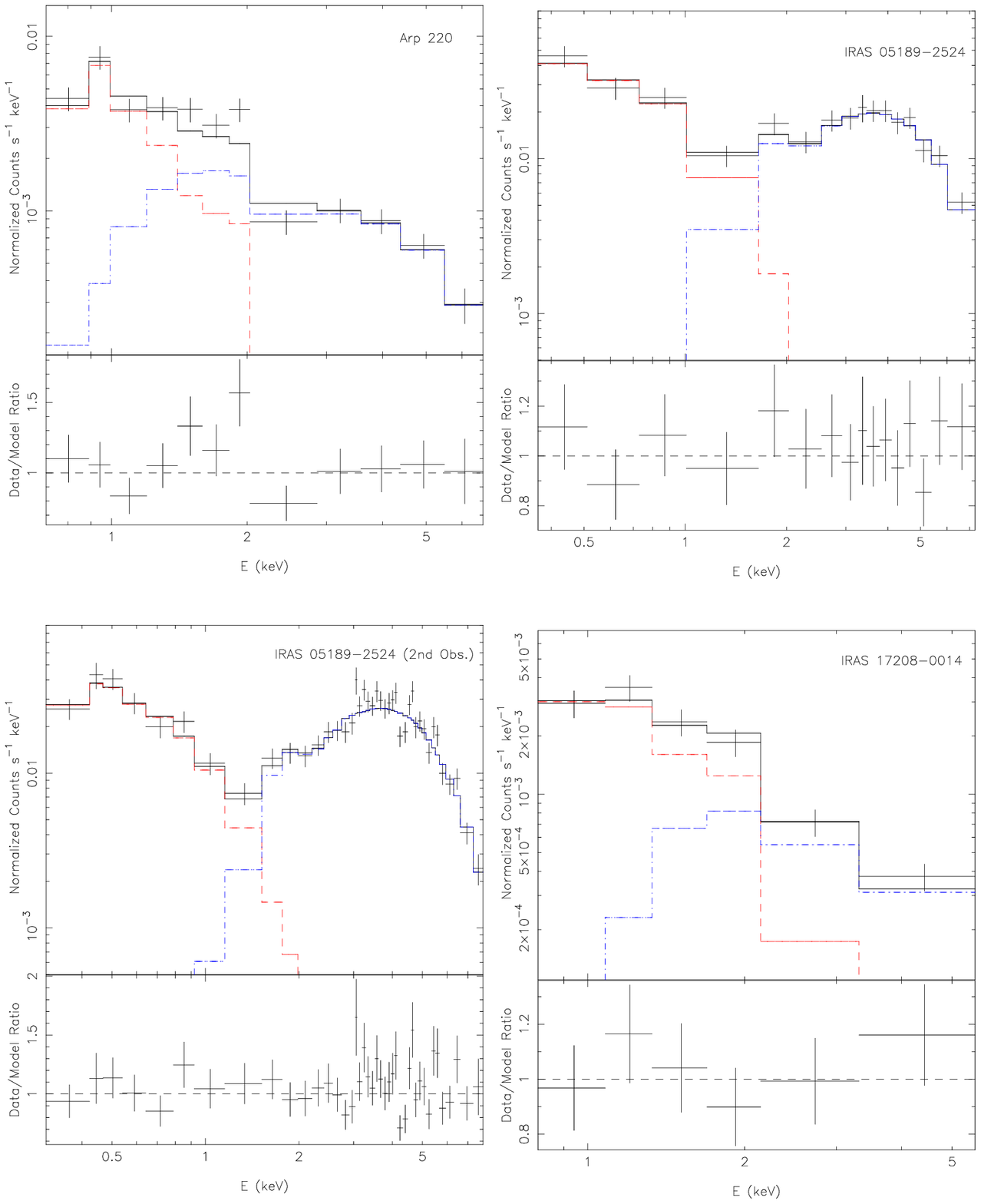}
\caption{}
\end{figure*}
\setcounter{figure}{2}
\begin{figure*}
\plotone{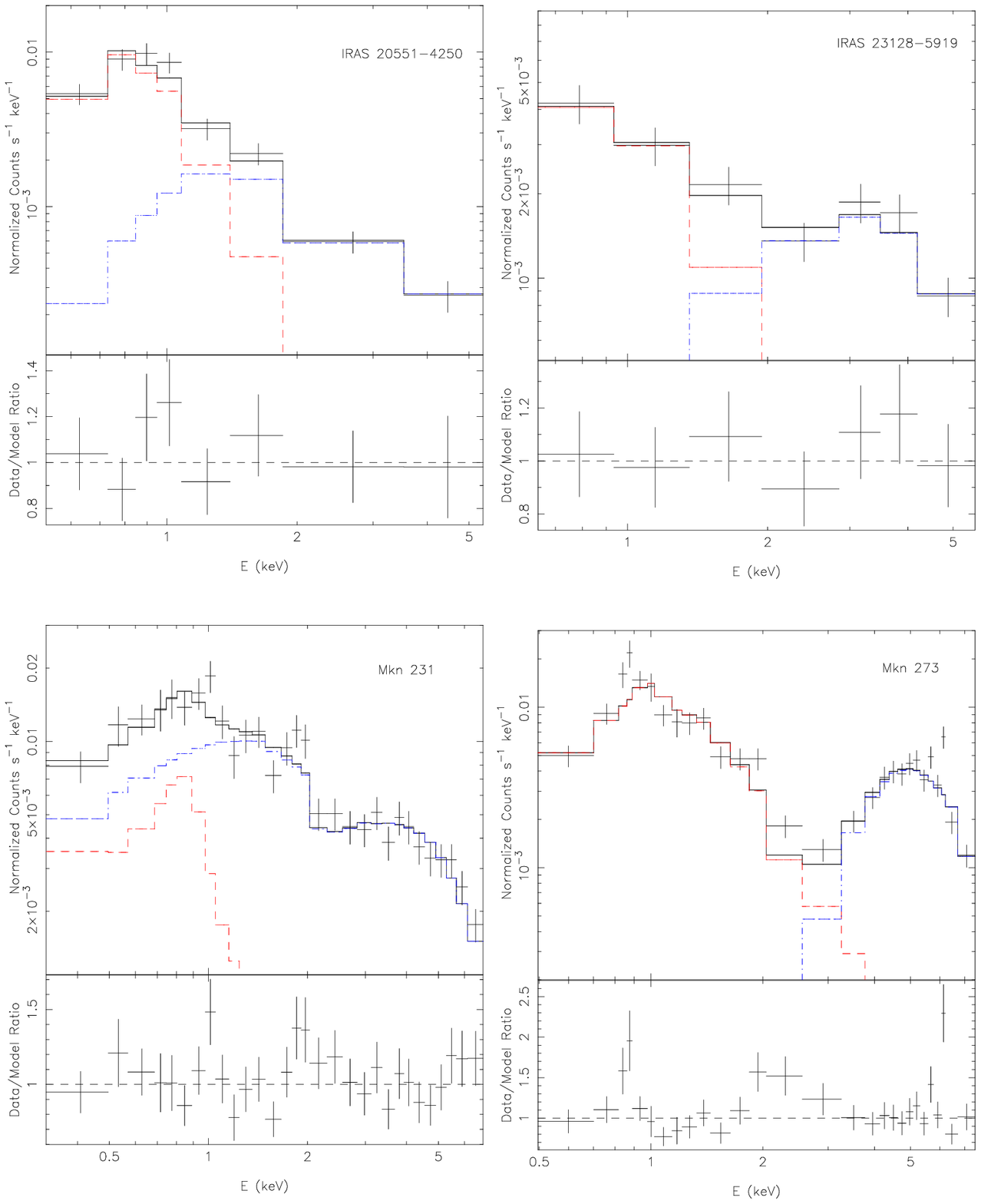}
\caption{(cont.)}
\end{figure*}
\setcounter{figure}{2}
\begin{figure*}
\plottwo{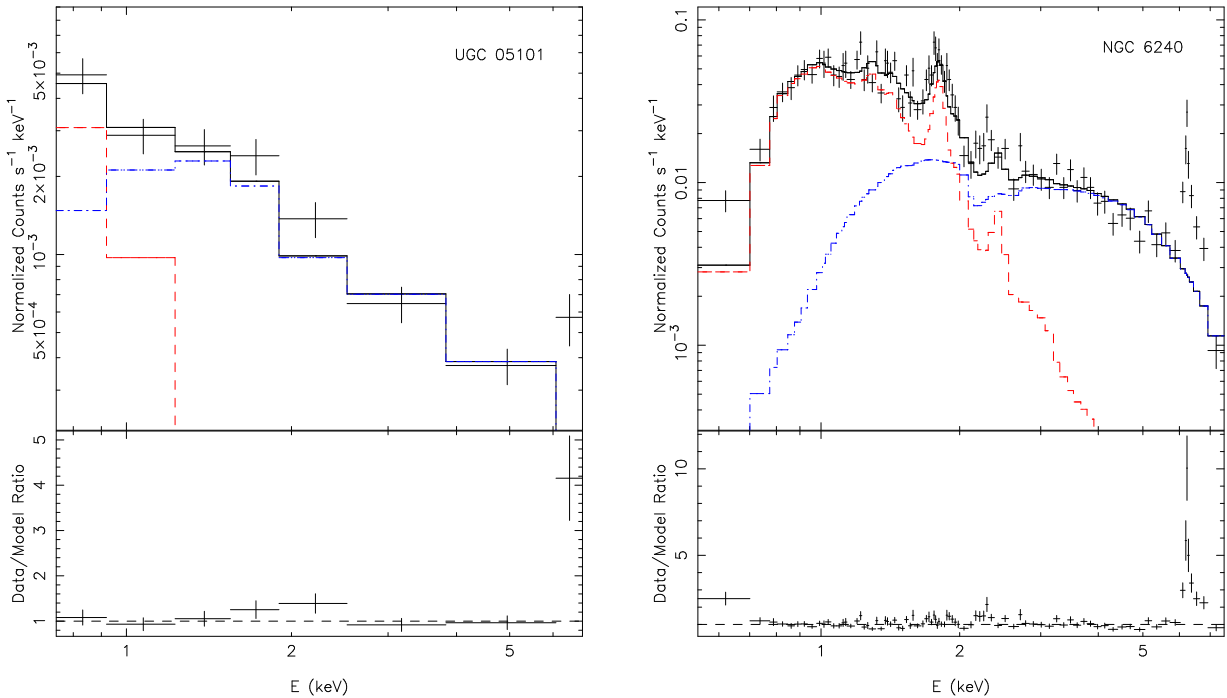}{f3j.eps}
\caption{Plasma + power-law fits to the nuclear spectra of the ULIRG galaxies.
The top panel shows the data and model and the data/model ratio is
shown in the bottom panel.  The plasma component model is shown with a
dashed red line and the power-law component model is shown with a
dot-dashed blue line.}
\end{figure*}
%\clearpage

\begin{deluxetable*}{llllllll}
\tablecaption{Plasma + Power-law Fits to Nuclear Spectra}
\tabletypesize{\scriptsize}
\tablehead{
\colhead{Galaxy} & \colhead{$N_H$} &
\colhead{kT (keV)} & \colhead{$Z/Z_{\odot}$} 
%& \colhead{K\tablenotemark{**} $\times 10^{5}$}
& \colhead{K\tablenotemark{**}}
& \colhead{$N_{H,2}$} &
\colhead{$\Gamma$} & \colhead{$\chi^2$/dof}\\
& \colhead{$(10^{22} \ \rm cm^{-2})$}
& \colhead{(keV)} 
& & & \colhead{$(10^{22} \ \rm cm^{-2})$}
}
\startdata
Arp 220 & 0.6 ($<1.1$) & 0.8 (0.6-1.1)& 2.1 ($>$0.2) &
1.3 (0.1 - 16) &
0.3 ($<$2.5) & 1.1 (0.3-1.8) & 19.5/17 \\
IRAS 05189-2524 & 0.01 ($<0.3$) & 0.7 ($<$1.3) & 0.01 ($<0.05$) &
24. (15.-250) & 
3.8 (2.2-6.0) & 1.0 (0.4-1.7) & 22.0/24 \\
IRAS 05189-2524* & 0.06 ($<0.2$) & 0.4 ($<$0.8) & 0.00 ($<0.03$) &
43. (16.-140) &
4.5 (3.4-5.8) & 1.1 (0.7-1.5) & 87/74 \\
IRAS 17208-0014 & 1.0 ($<1.5$) & 0.7(f) & 1.0(f) & 
4.5 ($<$13.) & 
0.6 ($<$39) & 1.9 ($>$ 0.5) &
6.8/7\\
IRAS 20551-4250 & 0.6 ($<$0.8) & 0.3 ($<1.0$) & 1.0(f) &
8.2 (0.6-21.) &
0.0 ($<$4.0) &
1.6 (0.8-3.8) & 11.2/9 \\
IRAS 23128-5919 & 0.2 ($<0.5$) & 0.7(f) & 0.0 ($<0.1$) & 
9.0 (2.5-18) & 4.8 (0.9-13.2) & 2.2 (0.5-4.6) & 8.0/8\\
Mkn 231 & 0.0 $<0.6$) & 0.7(f) & 0.15 ($0.02-3.2$) & 
1.8 (0.2-16) &
0.0 ($<0.6$) & 0.2 (0.0-0.5) & 41.6/50\\
Mkn 273 & 0.3 & 1.3 & 0.06 & 17 & 24 & 0.8 & 72.6/45\\
UGC 05101 & 0.2 & 0.4 & 1.0(f) & 5.6 & 0.3 & 1.2 & 29.3/9\\
NGC 6240 & 1.15 &  0.65  & 0.58 & 300. &
0. & 0.80 & 321/132 \\% n6240_nucl_cl_mekalpl.log
\enddata
\tablenotetext{*}{2nd observation}
\tablenotetext{**}{Plasma model normalization in units of 
$10^{-9}\frac{\int n_e n_H dV}{4 \pi D^2}$, $n_e$ = electron density
  in cm$^{-3}$, 
$n_p$ = Hydrogen density in cm$^{-3}$, D = luminosity distance to
  source in cm.}
\tablecomments{Errors were determined using $\Delta\chi^2=4.605$ for 
fits with $\chi^2/dof < 1.5$.  Parameters marked with ``f'' were
frozen at the value shown.  }
\end{deluxetable*}
%\clearpage

\subsection{Fe-K$\alpha$ Emission}
In Mkn 273 and NGC 6240, strong Fe-K emission is present, and Fe-K
emission is marginally detected in UGC 05101 (joint analysis of the  
Chandra and XMM-Newton data, to be presented elsewhere,  verifies the
existence of the line). In Table 6 we show the results of including a
Gaussian line in the plasma + power-law model fits for IRAS 05189-2524, Mkn
231, Mkn 273, NGC 6240, and UGC 05101 (insufficient counts are detected from
the other ULIRGs for these fits to result in useful constraints).  We
also calculate the upper limits for a narrow line (physical width =
0.01 keV; the resolution of ACIS-S at $\sim 6$ keV is $\sim 0.1$ keV),
and lines with physical widths of 0.1 and 0.5 keV at 6.7 keV using
absorbed power-law models applied to the 2-8 keV spectra of IRAS 05189-2524
and Mkn 231.  6.7 keV would correspond to the K$\alpha$ emission from
Fe XXV.  Such a line would be expected from an ionized
scattering region, and we computed the upper limits assuming a
non-negligible physical width to demonstrate the impact of a range of
ionization states being present.  
Note that the Fe-K line from IRAS 05189-2524 reported in \citet{se01} is not
detected in our Chandra (or XMM-Newton) data, although the
significance of the line in the prior observations was low and our
upper-limits are consistent with the errors.  

%\clearpage
\begin{deluxetable*}{llllllll}
\tablecaption{Fe-K Fits to Nuclear Spectra\label{fekfits}}
\tabletypesize{\scriptsize}
\tablehead{
\colhead{Galaxy} & \colhead{Bandpass} & \colhead{$N_H$} &
\colhead{$\Gamma$} & \colhead{Line E} & \colhead{Line $\sigma$} &
\colhead{Line EW} & \colhead{$\chi^2$/dof}\\
& \colhead{(keV)} & \colhead{$10^{22} \rm \ cm^{-2}$} & &
\colhead{(keV)} & \colhead{(keV)} & \colhead{(keV)}
}
\startdata
IRAS 05189\tablenotemark{*} & 0.4-8.0 & 4.5 (3.4-5.8) & 1.1 (0.7-1.5)
& 6.4f & 0.01f & 0.0 ($<$0.1) & 87/73 \\ % iras05189-2524_obs2_nucl_cl_mekalplga.log
& 2.0-8.0 & 4.4 (2.4-6.7) & 1.1 (0.6-1.7) & 6.4f & 0.01f & 0.0
($<$0.1) & 73/57 \\ % iras05189-2524_obs2_fek2_s0.01_e6.4.log
& 2.0-8.0 & 4.8 (2.7-7.2) & 1.3 (0.7-1.9) & 6.7f & 0.1f & 0.1 ($<$0.4)
& 71/57 \\ % iras05189-2524_obs2_fek2_s0.1_e6.7.log
& 2.0-8.0 & 4.7 (2.5-7.6) & 1.2 (0.6-2.1) & 6.7f & 0.5f & 0.1 ($<$0.7) 
& 73/57 \\ % iras05189-2524_obs2_fek2_s0.5_e6.7.log\\
Mkn 231 & 0.4-8.0 & 0 ($<$0.8) & 0.2 (0.0-0.7) & 6.4f & 0.01f & 0.2
($<$0.5) & 38/49 \\  % mkn231_nucl_cl_mekalplga.log
& 2.0-8.0 & 0.0 ($<$1.6) & 0.3 (0.0-0.9) & 6.4f & 0.01f & 0.2 ($<$0.6)
& 12/24 \\ % mkn231_fek2_s0.01_e6.4.log
& 2.0-8.0 & 0.0 ($<$1.5) & 0.3 (-0.1-0.9) & 6.7f & 0.1f & 0.1 ($<$0.5)
& 16/24 \\ % mkn231_fek2_s0.1_e6.7.log
& 2.0-8.0 & 0.0 ($<$2.3) & 0.5 (0.0-1.3) & 6.7f & 0.5f & 0.6 ($<$1.5)
& 13/24 \\ % mkn231_fek2_s0.5_e6.7.log
Mkn 273 & 0.4-8.0 & 24 (9-40) & 0.9 (-0.5-2.6) & 6.35 (6.30-6.39) & 
0.01 ($<$0.08) & 0.24 (0.09-0.44) & 61/42 \\ % mkn273_nucl_cl_mekalplga.log
UGC 05101 & 0.4-8.0 & 1.0 ($<$6.5) & 1.9 (0.80-4.7) & 6.32 (6.27-6.44)
& 0.01f & 5.9 (2.7-630) & 15/8 \\ % ugc05101_nucl_cl_mekalpl_solar_t8e6_ga.log
NGC 6240 & 0.4-8.0 & 0.4 ($<$1.9) & 1.4 (0.9-1.9) & 6.46 (6.39-6.57) &
0.20 (0.10-0.36) & 2.4 (1.7-3.2) & 217/129 \\ % n6240_nucl_cl_mekalplga.log
\enddata
\tablenotetext{*}{Second observation} 
\tablecomments{Fits to the 2.0-8.0 keV bandpass are based on the model
  given in Table 5 with the addition of a Gaussian component with the
  parameters shown here for the second (i.e., intrinsic) absorber,
  power-law slope and Gaussian line.  Fits to the 3.0-8.0 keV bandpass
  are based on an absorbed power-law plus Gaussian line model.}
\end{deluxetable*}
%\clearpage

The hard X-ray continuum is usually very weak, and in many cases the
resulting uncertainties on the continuum fit are substantial. Thus, a
more direct (model-independent) upper limit to the Fe-K$\alpha$ flux
was determined by interpolating the local background plus continuum
flux across the region of the Fe-K$\alpha$ line in the spectra, and
then setting an upper bound to the excess of line photons.  As a
consistency check, we computed the flux of the Fe-K line in Mkn 273
directly from the spectrum in this fashion, which resulted in a value
of $4.5 \times 10^{-14}$ \ergcms which is consistent with the value of
$6.0 \times 10^{-14}$ \ergcms obtained from the fit listed in Table 6.
The resulting upper limits on the K$\alpha$ 
fluxes are listed in Table 7, where line fluxes derived from spectral
fits are given for the galaxies listed in Table 6 (i.e., those for which there
were sufficient counts for spectral fitting in the Fe-K region). Note
that the lines are typically 
about two orders-of-magnitude smaller than the flux of the strong
K$\alpha$ line in NGC 6240.

Because the hard X-ray continua are so weak, the upper limits on the
equivalent widths (EW) of the K$\alpha$ lines implied by the above are
usually of-order a keV. The spectra of IRAS 05189-2524 and
Mkn 231 have equivalent width 
upper limits of 0.1 keV and 0.5 respectively for a narrow K$\alpha$ line
($\sigma=0.01$ keV) at 6.4 keV. Note that in
the cases of UGC 05101, IRAS 17208-0014 and IRAS 20551-4250,
insufficient counts exist in the 3-8 keV bandpass to even fit the power-law
plus Gaussian model.

In general, obscured
(narrow-line) AGN in which 
the hard X-ray emission is dominated by reflection/reprocessing tend to have
lines with EWs in excess of 1 keV (e.g., \citet{p96} and 
references therein), and similarly hot gas is expected to   
have Fe-K$\alpha$ emission at 6.7 keV with an equivalent width of 
$\sim 0.6 (Z/Z_{\odot})$ keV  \citep{ra85}. Thus, in most cases the
upper bounds on the EW are not terribly restrictive. The upper limits
on the K$\alpha$ fluxes are more constraining in these cases (as we
will discuss below). 

%In Mkn 231, we find that the addition on an
 %narrow line (for 2 additional model
%parameters) reduces $\chi^2$ by 4 with a best-fit energy of 6.4 keV and an
%EW of 0.39 (0.08-0.70) keV.  

%\clearpage
\begin{deluxetable}{lll}
\tablecaption{Chandra Fe-K Line Limits}
\tabletypesize{\scriptsize}
\tablehead{
\colhead{Galaxy} & \colhead{$F_{\rm Fe-K}\ \rm (10^{-14}\ ergs\ s^{-1}$)} &
\colhead{Method}
}
\startdata
Arp 220 & $<$0.92 & Direct measurement\\
%IRAS 05189 & $<$7.1 & This work\\
IRAS 05189 & $<$5.3 & Spectral fit\\ % iras05189-2524_obs2_nucl_cl_mekalplga.log
IRAS 17208 & $<$0.83 & Direct measurement\\
IRAS 20551 & $<$0.88 & Direct measurement\\
IRAS 23128 & $<$1.3 & Direct measurement\\
Mkn 231 & $<$5.0 & Spectral fit\\ % mkn231_fek_s0.01_e6.4.log
Mkn 273 & 6.0 (2.2-11)  & Spectral fit\\ %
% mkn273_nucl_cl_mekalplga.log
UGC 05101 & 3.5 (1.6-380) & Spectral fit\\ % ugc05101_nucl_cl_mekalpl_solar_t8e6_ga.log
NGC 6240 %$29 \pm 7.0$ & \cite{iw98} \\ % 
& 30. (22.-40.) & Spectral fit\\ % n6240_nucl_cl_mekalplga.log
\enddata
\tablenotetext{*}{Method used to determine the Fe-K flux, as described
  in the text.  Briefly, for sources with insufficient counts for
  spectral fitting, ``direct measurement'' refers to determining an
  upper-limit to the Fe-K flux based on the background level at 6.4
  keV in the rest frame.}
\end{deluxetable}
%\clearpage

% ---- The following two sections are new
\subsection{More Complex Models}
The reduced $\chi^2$ values of plasma + power-law + Fe-K line fits to
Mkn 273, NGC 6240, and UGC 05101 are $\ga 1.5$.  In the case of UGC
05101 the fit is nevertheless statistically acceptable due to the low
number of degrees of freedom.  We added an addition plasma model
(mekal) model to the Mkn 273 and NGC 6240 fits.  In the case of Mkn
273 this results in an acceptable fit with $\chi^2/dof = 41.2/40$, however the 
photon index of the power-law component is poorly constrained $\Gamma
= 1.7 \ (0.3-3.8)$.   Since a motivation for these spectral fits is to
compute accurate fluxes for comparison with global (i.e., spatially-averaged)
spectral fits (see below), we fixed the photon index at 1.8, and the
results of this fit are shown in Table \ref{mkn273fits}.  For NGC 6240
a more complex model was necessary in order to fit both the nuclear
and global spectrum.  This model, motivated by the
BeppoSAX fit given in \citet{vi99}, contains two plasma components,
a power-law (representing AGN emission scattered by a
scattering region), an Fe-K complex (modeled with multiple Gaussian
components) and a Compton reflection component \citep{ma95}.  
See Table \ref{n6240fits} for the fit parameters.
We also found it necessary to include an
additional Gaussian component with an observed (rest-frame) energy of
$\sim$ 2.2 ($\sim 2.3$) keV, however note that the observed energy is
close to the energies of significant absorption edges in the Chandra
mirror response where the calibration tends to be the most uncertain,
and accordingly this feature may not be real.
\begin{deluxetable}{lll}
\tablecaption{Double Plasma + Power-law Fits to Mkn 273
  Spectra\label{mkn273fits}} 
\tabletypesize{\scriptsize}
\tablehead{
\colhead{Parameter} &
\colhead{Nuclear Spectrum} &
\colhead{Global Spectrum}
}
\startdata
% Values here are from mkn273_nucl_cl_2mekalplga_gamma1.8_03jan03.log
% and mkn273_cl_2mekalplga_gamma1.8_07jan03.log
$N_{H,1} \ (10^{20} \rm \ cm^{-2})$ & 0.5 ($<1.8$) & 2.1 ($<$7.8)\\
$kT_1$ (keV) & 0.84 (0.68-0.97) & 0.66 (0.59-0.71)\\
$Z_1 \ (/Z_{\odot})$ & 1.0f & 0.15 (0.10-0.23)\\
$K_1$\tablenotemark{*}& 0.75 (0.50-1.4) & 17 (1.1-31) \\ % 10^5 K
$N_{H,2} \ (10^{22} \rm \ cm^{-2})$ & 1.2 (0.7-1.9) & 1.3 (0.9-2.2)\\
$kT_2$ (keV) & 1.7 (1.0-2.0)\tablenotemark{\ddag} & 1.1 (0.87-1.4)\\
$Z_2 \ (/Z_{\odot})$ & 1.0f & 0.9 (0.1-5.0)\tablenotemark{\ddag}\\
$K_2$\tablenotemark{*}& 12 (8.4-20) & 22 (4.5-64)\\
$N_{H,3} \ (10^{23} \rm \ cm^{-2})$ & 3.5 (3.0-4.3) & 3.4 (2.7-4.2)\\
Line E (keV) & 6.35 (6.30-6.39) & 6.37 (6.30-6.42)\\
Line $\sigma$ (kev) & 0 ($<0.08$) & 0.0 ($<0.10$)\\
Line N\tablenotemark{$\dagger$} & 7.2 (2.6-11.8) & 7.2 (2.2-13) \\ % 10^6 N
$\Gamma$ & 1.8f & 1.8f\\
$N \ (\times{10^4})$ & 7.4 (5.9-9.6) & 7.6 (5.7-10) \\
%$F_{0.5-2.0 \rm \ keV}$\tablenotemark{3} & 0.34 & 1.1 \\
%$F_{2.0-10.0 \rm \ keV}$\tablenotemark{3} & 7.9 & 8.3 \\
$\chi^2/dof$ & 41.2/41 & 118/103 \\
\enddata
\tablecomments{The model fitted to the spectra was: $Abs(N_{H,1}) [
  Mekal(kT_1, Z_1, K_1) + Abs(N_{H,2})\times Mekal(kT_2, Z_2, K_2) +
  Abs(N_{H,3})\times (PL(\Gamma, N) + Gaussian(E, \sigma,
  N_{Line}))]$, with Abs = absorption, Mekal=plasma, PL = power law
  and Gaussian = Gaussian components.}
\tablenotetext{*}{Plasma normalization in units of $10^{-9}\frac{\int
  n_e n_H dV}{4 \pi D^2}$, $n_e$ = electron density in cm$^{-3}$,  
  $n_p$ = Hydrogen density in cm$^{-3}$, D = luminosity distance to
  source in cm.}
 \tablenotetext{$\dagger$}{Gaussian line normalization in units of $10^{-6}$
  photons cm$^{-2}$ s$^{-1}$.}
 \tablenotetext{\ddag}{Parameter reached pre-set boundary during error search.}
\end{deluxetable}

\begin{deluxetable}{lll}
\tablecaption{Complex Spectral Fits to NGC 6240
  Spectra\label{n6240fits}} 
\tabletypesize{\scriptsize}
\tablehead{
\colhead{Parameter} &
\colhead{Nuclear Spectrum} &
\colhead{Global Spectrum}
}
\startdata
% Data is from n6240_nucl_cl_2mekal2pl3gapexrav_30dec.log 
% and n6240_cl_2mekal2pl3gapexrav_30dec.log
% Revised 1/28/03 to be from n6240_nucl_cl_2mekalpl3gapexrav_07jan03.log
% and 
$N_{H,1} \ (10^{21} \rm \ cm^{-2})$ & 0.4 ($<$5.5) & 0.5 (0.3-0.9)\\
$N_{H,2} \ (10^{21} \rm \ cm^{-2})$ & 5.4 ($<$9.4) & 0.3 ($<$1.5)\\
$kT_1$ (keV) & 0.23 (0.17-0.38) & 0.30 (0.27-0.33)\\
$Z_1 \ (/Z_{\odot})$ & 2.6 (0.1-5.0)\tablenotemark{\ddag} & 0.32 (0.11-1.1)\\
%$K_1$\tablenotemark{*} ($\times 10^{3}$)& 0.051 (0.004-0.19) & 0.4
%  (0.1-14) \\ % 10^3 K
$K_1$\tablenotemark{*}  & 0.51 (0.04-1.9) & 4.1
  (1.3-13.) \\ % 10^3 K
$N_{H,3} \ (10^{22} \rm \ cm^{-2})$ & 1.3 (0.77-1.4) & 0.78 (0.69-0.90)
  \\
$kT_2$ (keV) & 0.67 (0.61-0.76) & 0.66 (0.62-0.72)\\
$Z_2 \ (/Z_{\odot})$ & 9.0 (3.1-10)\tablenotemark{\ddag} & 0.76 (0.32-1.23)\\
%$K_2$\tablenotemark{*} ($\times 10^{4}$)& 1.3 (0.8-3.8) & 21 (5.1-29)\\
$K_2$\tablenotemark{*} & 1.3 (0.88-3.9) & 21 (3.5-41.)\\
Line1 E (keV) & 2.31 (2.25-2.35) & 2.33 (2.31-2.35)\\
Line1 $\sigma$ (kev) & 0.01f & 0.01f\\
Line1 N\tablenotemark{$\dagger$}& 5.4 (2.0-9.0) & 12. (7.2-16) \\ % 10^6 N
$N_{H,4} \ (10^{22} \rm \ cm^{-2})$ & 1.5 (0.5-2.3)
  & 1.1 (0.7-1.7) \\
$Gamma$ & 1.55 (0.98-2.02) & 1.62 (1.22-1.93)\\
$N \ (\times{10^4})$ & 1.8 (0.9-3.5) & 2.7 (1.3-4.2)\\
Line2 E (keV) & 6.35 (6.33-6.38) & 6.34 (6.31-6.37)\\
Line2 $\sigma$ (kev) & 0.00 ($<$0.06) & 0.02 ($<$0.08)\\
Line2 N\tablenotemark{$\dagger$}& 16 (11-21) & 17 (11-24) \\ % 10^6 N
Line3 E (keV) & 6.57 (6.50-6.69) & 6.62 (6.54-6.67) \\
Line3 $\sigma$ (kev) & 0.01 ($<0.20$) & 0.01 ($<$0.13)\\
Line3 N\tablenotemark{$\dagger$}&  4.9 (0.8-11) & 7.4 (2.2-13)\\ % 10^6 N
$N_{H,5} \ (10^{24} \rm \ cm^{-2})$ & 1.3 (0.5-3.0) & 1.3 (0.7-2.3)\\
$N_{refl}$ & 2.5 (0.5-5.7) & 4.6 (1.4-13)\\ % 10^2 N
$\chi^2/dof$ & 147/117 & 331/225 \\ 
\enddata
\tablecomments{The model fitted to the spectra was: $Abs(N_{H,1})
  \times [
  Abs(N_{H,2}) \times Mekal(kT_1, Z_1, K_1) + Abs(N_{H,2}) \times 
  [Mekal(kT_2, Z_2, K_2) + Gaussian(E_1, \sigma_1, N_{Line,1}) +
  Abs(N_{H,3})\times [PL(\Gamma, N) + Gaussian(E_2, \sigma_2, N_{Line,2}) +
  Gaussian(E_3, \sigma_3, N_{Line,3})
  ] + Abs(N_{H,4}) \times Refl(N_{refl})]$, with Abs = absorption,
  Mekal=plasma, PL = power law, Gaussian = Gaussian components and
  Refl = Compton reflection components.}
\tablenotetext{*}{Plasma normalization in units of $10^{-10}\frac{\int
  n_e n_H dV}{4 \pi D^2}$, $n_e$ = electron density in cm$^{-3}$,  
  $n_p$ = Hydrogen density in cm$^{-3}$, D = luminosity distance to
  source in cm.}
 \tablenotetext{$\dagger$}{Gaussian line normalization in units of $10^{-6}$
  photons cm$^{-2}$ s$^{-1}$.}
 \tablenotetext{\ddag}{Parameter reached pre-set boundary during error search.}
\end{deluxetable}

\subsection{Spatially-Averaged Spectra}
In order to determine the amount of flux extended beyond the nuclear
region we fitted the spectra extracted from the ``global'' regions
shown in Figure 1 with a models consisting of a plasma plus absorbed
power-law (i.e., as fit to the nuclear spectra) and two plasmas and an absorbed
power-law.  The parameters from the statistically-preferred model
(based on the f-test statistic) are given in Table \ref{globalfits}.
The spectra from Mkn 273 and NGC 6240 were fit with the same models
used for the nuclear spectra, with the results shown in Tables
\ref{mkn273fits} and \ref{n6240fits}.

%%\begin{deluxetable}{llllllllll}
%%\rotate
%%\tablecaption{Double Plasma + Power-law Fits to Global Spectra}
%%\tabletypesize{\scriptsize}
%%\tablehead{
%%\colhead{Parameter} & \colhead{Arp 220} & \colhead{I05189} & 
%%\colhead{I17280} & \colhead{I20551} & \colhead{I23128} & 
%%\colhead{Mkn 231} & \colhead{Mkn 273} & \colhead{UGC 05101} &
%%\colhead{NGC 6240}
%%}
%%\startdata
%%$\Gamma$ &
%%$1.14_{-0.87}^{+0.94}$ &
%%$1.14_{-0.87}^{+0.94}$ &
%%$1.14_{-0.87}^{+0.94}$ &
%%$1.14_{-0.87}^{+0.94}$ &
%%$1.14_{-0.87}^{+0.94}$ &
%%$1.14_{-0.87}^{+0.94}$ &
%%$1.14_{-0.87}^{+0.94}$ &
%%$1.14_{-0.87}^{+0.94}$ &
%%$1.14_{-0.87}^{+0.94}$
%%\enddata
%%\end{deluxetable}
%%%\clearpage
%%

%%%% \input{global_fits_table}
\begin{deluxetable*}{llllllllll}
\tablecaption{Fits to Global Spectra\label{globalfits}}
\tabletypesize{\scriptsize}
\tablehead{
\colhead{Parameter}
& \colhead{Arp 220}
& \colhead{I05189}
& \colhead{I17208}
& \colhead{I20551}
& \colhead{I23128}
& \colhead{Mkn 231}
& \colhead{UGC 05101}
}
\startdata
$N_{H,1}\ (10^{21} \rm \ cm^{-2})$
& $0.20_{-0.00}^{+0.94}$
 \tablenotemark{\ddag}& $1.39_{-1.25}^{+0.48}$
& $5.22_{-4.72}^{+4.02}$
& $0.20_{-0.00}^{+2.82}$
 \tablenotemark{\ddag}& $2.98_{-1.11}^{+2.17}$
& $0.05_{-0.00}^{+0.79}$
 \tablenotemark{\ddag}& $2.29_{-2.14}^{+3.12}$
\\
$kT_1$ (keV)
& $0.33_{-0.03}^{+0.06}$
& $0.30_{-0.00}^{+0.31}$
 \tablenotemark{\ddag}& $0.35_{-0.05}^{+0.96}$
& $0.32_{-0.02}^{+0.12}$
& $0.50_{-0.20}^{+0.19}$
& $0.53_{-0.23}^{+0.15}$
& $0.70_{-0.40}^{+0.34}$
\\
$Z_1\ (/Z_{\odot})$
& 1.0f
& $0.0_{-0.0}^{+0.0}$
 \tablenotemark{\ddag}& $0.1_{-0.1}^{+4.9}$
& 1.0f
& $0.1_{-0.0}^{+0.1}$
& $0.2_{-0.1}^{+0.3}$
& $0.1_{-0.0}^{+0.1}$
\\
%$K_1$\tablenotemark{*} ($\times 10^{5}$)
$K_1$\tablenotemark{*}
& 1.20
& 141.40
& 
& 
& 
& 
& 
\\
$N_{H,2}\ (10^{22} \rm \ cm^{-2})$
& $0.7_{-0.5}^{+0.2}$
& $10.6_{-8.8}^{+89.4}$
& 
& 
& 
& 
& 
\\
$kT_2$ (keV)
& $0.8_{-0.2}^{+0.1}$
& $2.0_{-0.9}^{+0.0}$
 \tablenotemark{\ddag}& 
& 
& 
& 
& 
\\
$Z_2\ (/Z_{\odot})$
& 1.0f
& $0.6_{-0.5}^{+4.4}$
& 
& 
& 
& 
& 
\\
%$K_2$\tablenotemark{*} ($\times 10^{5}$)
$K_2$\tablenotemark{*}
& 8.84
& 563.70
& 
& 
& 
& 
& 
\\
$N_{H,3}\ (10^{22} \rm \ cm^{-2})$
& $1.0_{-1.0}^{+2.9}$
& $2.8_{-0.5}^{+1.1}$
& $0.0_{-0.0}^{+6.2}$
 \tablenotemark{\ddag}& $0.7_{-0.6}^{+0.3}$
& $5.1_{-3.9}^{+6.8}$
& $0.0_{-0.0}^{+2.1}$
 \tablenotemark{\ddag}& $0.7_{-0.7}^{+4.6}$
\\
$\Gamma$
& $1.14_{-0.87}^{+0.93}$
& $0.52_{-0.23}^{+0.55}$
& $1.68_{-1.22}^{+1.69}$
& $0.85_{-1.61}^{+2.04}$
& $2.59_{-1.31}^{+1.95}$
& $0.55_{-0.41}^{+0.37}$
& $1.06_{-1.01}^{+1.79}$
\\
$N_3$ ($\times 10^{5}$)
& 1.38
& 11.32
& 1.53
& 0.51
& 21.59
& 2.85
& 0.80
\\
Line E (keV)
& 
& 
& 
& 
& 
& 
& $6.90_{-0.29}^{+0.00}$
\\
Line $\sigma$ (keV)
& 
& 
& 
& 
& 
& 
& $0.37_{-0.37}^{+0.13}$
\\
Line N\tablenotemark{$\dagger$} ($\times 10^{6}$)
& 
& 
& 
& 
& 
& 
& $4.53_{-2.91}^{+68.60}$
\\
$\chi^2/dof$& 48.9/64& 94.4/75& 18.7/16& 21.0/27& 30.7/30& 83.1/89& 32.5/16\\

\enddata
%\tablecomments{The model fitted to the spectra was the same : $Abs(N_{H,1}) [
%  Mekal(kT_1, Z_1, K_1) + Abs(N_{H,2})\times Mekal(kT_2, Z_2, K_2) +
%  Abs(N_{H,3})\times (PL(\Gamma, N) + Gaussian(E, \sigma,
%  N_{Line}))]$, with Abs = absorption, Mekal=plasma, PL = power law
%  and Gaussian = Gaussian components.}
\tablenotetext{*}{Plasma normalization in units of $10^{-9}\frac{\int
  n_e n_H dV}{4 \pi D^2}$, $n_e$ = electron density in cm$^{-3}$,  
  $n_p$ = Hydrogen density in cm$^{-3}$, D = luminosity distance to
  source in cm.}
 \tablenotetext{$\dagger$}{Gaussian line normalization in units of $10^{-6}$
  photons cm$^{-2}$ s$^{-1}$.}
 \tablenotetext{\ddag}{Parameter reached pre-set boundary during error search.}
\end{deluxetable*} 

\subsection{Continuum Fluxes and Luminosities}
The observed fluxes and luminosities in the 0.5-2.0 keV and 2.0-10.0 keV
bandpasses are given in Table \ref{fluxtable},
derived from the plasma + power-law
fits to the nuclear spectra, except for Mkn 273, UGC 05101 and NGC
6240 where the fit in Table 6 was used.  We also include the fluxes derived
from fits to the ``global'' spectra (using the models shown in Tables
\ref{mkn273fits} - \ref{globalfits}), from which it can be seen that
on average
$\sim 50\%$ and 75\% of the 0.5-2.0 keV and 2.0-10.0 keV total flux originates
in the $\sim 0.5-1.0$ kpc scale nuclear regions.   Since the sources with
higher signal-to-noise require more complex models to fit their spectra it is
likely that most or all ULIRGs are similarly complex 
\citep[see, e.g., XMM-Newton spectral fits in][]{br02}. Fitting the spectra
with models that are not sufficiently complex physically but are acceptable
statistically may result in incorrect flux estimates, especially in the case
of the 2-10 keV flux estimates when the power-law slope is poorly constrained.
In order to derive a model-independent estimate of the amount of extra-nuclear
flux we also list in Table \ref{fluxtable} the ratio of the nuclear and global
count rates in the 0.5-2.0 and 2.0-8.0 keV bandpasses (with the 2.0-8.0 keV
count rate ratios being listed with the 2-10 keV flux ratios).   As expected
the flux and count rates ratio are in fairly good agreement, but differ by up
to $\sim 20\%$.

%\clearpage
% --- Values below have been revised
\begin{deluxetable*}{l|lll|ll|llll}
%\rotate
\tablecaption{Fluxes and Luminosities\label{fluxtable}}
\tabletypesize{\scriptsize}
\tablehead{
\colhead{Galaxy} & \multicolumn{3}{c}{Global} & \multicolumn{2}{c}{Nuclear} &
%\colhead{$F_{\rm Nuclear,\ 0.5-2.0 keV}$/$F_{\rm Global,\ 0.5-2.0 keV}$} &
%\colhead{$F_{\rm Nuclear,\ 0.5-2.0 keV}$/$F_{\rm Global,\ 0.5-2.0 keV}$} \\
\multicolumn{2}{c}{0.5-2.0 keV} & 
\multicolumn{2}{c}{2-10 keV}
\\
\colhead{} & 
%\colhead{$C_{0.5-2.0\rm \ keV}$} & 
%\colhead{$C_{2-10\rm \ keV}$} &
\colhead{$F_{0.5-2.0\rm \ keV}$} & 
\colhead{$F_{2-10\rm \ keV}$} &
\colhead{$L_{\rm 2-10 \ keV}$} &
%\colhead{$C_{0.5-2.0\rm \ keV}$} & \colhead{$C_{2-10\rm \ keV}$}&
\colhead{$F_{0.5-2.0\rm \ keV}$} & \colhead{$F_{2-10\rm \ keV}$}
& \colhead{$\frac{F_N}{F_G}$}
& \colhead{$\frac{C_N}{C_G}$}
& \colhead{$\frac{F_N}{F_G}$}
& \colhead{$\frac{C_N}{C_G}$} 
}
\startdata
Arp 220 & 0.6 & 1.3 & 0.10 & 0.16 & 1.2 & 0.27 & 0.25 & 0.85 & 0.65\\
IRAS 05189-2524* & 0.73 & 37. & 15. & 0.66 & 35. & 0.90 & 0.87 & 0.96 & 0.95\\
IRAS 17208-0014 & 0.21 & 0.62 & 0.25 & 0.11 & 0.42 & 0.52 & 0.43 & 0.68 & 0.77\\
IRAS 20551-4250 & 0.49 & 0.84 & 0.35 & 0.18 & 0.39 & 0.37 & 0.37 & 0.46 & 0.42 \\
IRAS 23128-5919 & 0.43 & 1.5 & 0.66 & 0.13 & 1.3 & 0.31 & 0.25 & 0.92 & 0.67\\
Mkn 231 & 1.0 & 7.8 & 3.1 & 0.53 & 8.1 & 0.53 & 0.47 & 1.04 & 0.85 \\
Mkn 273 & 1.1 & 8.3 & 2.6 & 0.34 & 7.8 & 0.30 & 0.26 & 0.95 & 0.88\\ 
UGC 05101 & 0.20 & 1.6 & 0.55 & 0.12 & 0.81 & 0.60 & 0.52 & 0.51 & 0.76\\
NGC 6240 & 7.2 & 26. & 3.4 & 1.8 & 17. & 0.26 & 0.23 & 0.66 & 0.63 \\
\enddata
\tablecomments{Fluxes are in units of $10^{-13}$ \ergcms, luminosities are in
units of $10^{42}$ \ergs, derived from plasma plus power-law fits.
The ``nuclear'' 2-10 keV flux exceeds the ``global'' flux in Mkn 231
due to statistical uncertainty in the power-law slope; the global 2-10
keV flux of Mkn 231 is 8.9 $\times 10^{-13}$ \ergcms when the
power-law slope is fixed at best-fit value of the nuclear spectrum fit
($\Gamma = 0.21$). $\frac{F_N}{F_G}$ is the ratio of nuclear and
global fluxes while $\frac{C_N}{C_G}$ is the ratio of nuclear and
global count rates.}
\end{deluxetable*} 
%\clearpage

\section{Variability}
Significant short-term variability in the 2.0-8.0 keV bandpass is not
observed in the events extracted from the nuclear regions when tested
by comparing the source and background photon arrival times with the
K-S statistic, with the exception of Mkn 231 \citep{ga02}.  In Figure
4 we plot the long-term 2-10 keV light curves of the ULIRGs for which 
ASCA and/or BeppoSAX data exist (listed in Table \ref{ascasaxtab}).  
The Chandra fluxes where derived 
from fits to the ``global'' regions shown in Figure 1, while the
typical ASCA and BeppoSAX source regions were 4-6'.  For the ASCA
observation of UGC 05101 and the two BeppoSAX observation of Arp 220,
the source was not 
detected with sufficient counts for spectral modeling so the count
rate was determined from the images and 2-10 keV fluxes were derived
assuming the best-fitting spectral model to the Chandra data.  While
there is some apparent variability in IRAS 17208-0014 and 
IRAS 20551-4250, we note
that the total flux of all sources detected within the central 4' of
the Chandra observations of these galaxies was 1.6$\times
10^{-13}$ \ergcms~ and 4.6$\times 10^{-13}$ \ergcms, respectively.
This suggests that some or all of the observed variability is due to
source confusion in the ASCA data.  
In summary, hard X-ray variability
is clearly observed in IRAS 05189-2524, Mkn 231, and Mkn 273 (the
AGN ULIRGs) and in NGC 6240 at levels of $\sim 40\%$ or greater, while in the
remaining ULIRGs no variability is observed that is not possibly due to
aperture effects.

%\clearpage
\begin{figure}[htbn]
\plotone{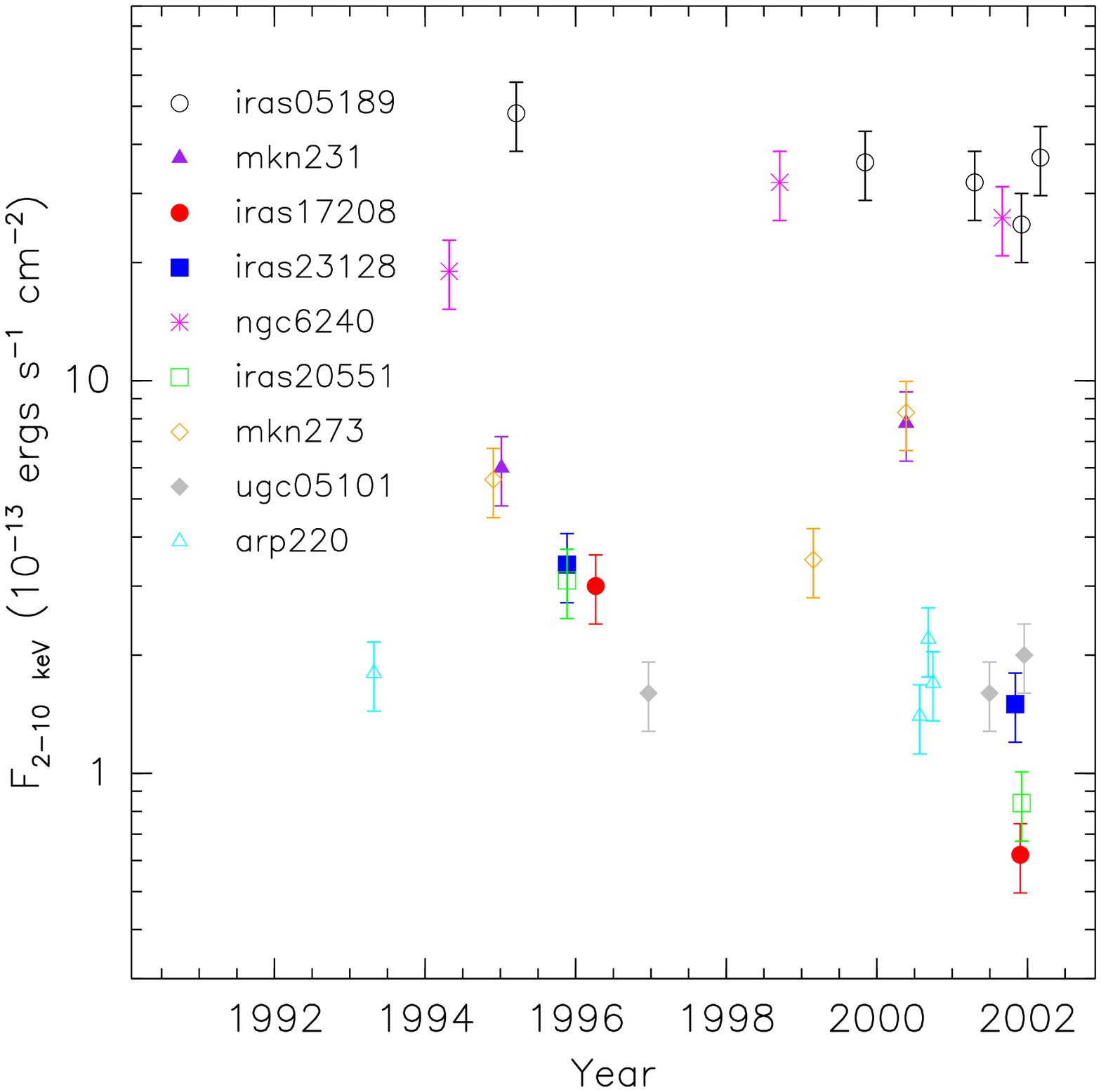}
\caption{Long-term light curves of ULIRG sample.}
\end{figure}

%\clearpage
\begin{deluxetable*}{lllllll}
\tablecaption{ASCA and BeppoSAX Fluxes and Fe-K Line Limits\label{ascasaxtab}}
\tabletypesize{\scriptsize}
\tablehead{
\colhead{Galaxy} & \colhead{Satellite} & \colhead{Date} &
\colhead{$F_{0.5-2.0 \ \rm keV}$} &
\colhead{$F_{2-10 \ \rm keV}$} & 
\colhead{Fe-K Line EW (keV)} & \colhead{References}
}
\startdata
Arp 220 & ASCA & 03/27/1994 & 0.6 & 1.8 & \nodata & 1\\
Arp 220 & BeppoSAX & 08/04/2000 & \nodata & 2.2 & \nodata & 7\\
Arp 220 & BeppoSAX & 08/27/2000 & \nodata & 1.7 & \nodata & 7\\
Mkn 231 & ASCA & 12/05/1994 & 1.2 & 6.0 & $<0.9$ & 1\\
Mkn 273 & ASCA & 10/27/1994 & 1.6 & 5.6 & 0.52 (0.26-1.0) & 1 \\
Mkn 273 & BeppoSAX & 01/25/1999 & \nodata & 3.5 & 1.2 (0.2-3.2) & 2\\
IRAS 05189-2524 & ASCA & 02/15/1995 & \nodata & 48 & 0.11 (0.03-0.18) & 4\\
IRAS 05189-2524 & BeppoSAX & 10/03/1999 & \nodata & 36 & 0.14 (0.05-0.34) & 4\\
IRAS 17208-0014 & ASCA & 03/06/1996 & \nodata & 3.0 & \nodata & 2\\
IRAS 20551-4250 & ASCA & 10/19/1995 & \nodata & 3.1 & \nodata & 3\\
IRAS 23128-5919 & ASCA & 02/11/1995 & \nodata & 3.4 & \nodata & 3\\
UGC 05101 & ASCA & 11/17/1996 & 1.7 & 6.2 & \nodata & 7\\
NGC 6240 & ASCA & 03/27/1994 & 6.4 & 19. & 2.2 & 5 \\
NGC 6240 & BeppoSAX & 08/15/1998 & 7.4 & 32. & 1.6 & 6\\
\enddata
\tablecomments{Fluxes in units of $10^{-13}$ \ergcms.}
\tablerefs{1. \citet{iw99}; 2. \citet{ri00}; 
3. \citet{mi99}; 4. \citet{se01}; 5. \citet{iw98}; 6. \citet{vi99};
7. This work 
}
\end{deluxetable*}
%\clearpage

\section{Discussion}

Our overall goal in this paper is to better understand the role of AGN
in the ULIRG phenomenon. We can cast this in terms of three related
questions: 1) Is the signature of an AGN present in the Chandra hard
X-ray data? 2) Does an AGN dominate the hard X-ray emission? 3) Do the
X-ray data imply that an obscured AGN dominates the bolometric energy
output? To address these questions, we will consider the
structure/morphology of the X-ray emission (information that Chandra
uniquely provides), the spectral properties of the emission, and the
ratio of the hard X-ray to bolometric (IR) luminosity. 

As noted in section 1, optical and IR spectroscopy of ULIRGs reach a
surprising degree of agreement as to their classification. ULIRGs that
are optically classified as HII-regions or LINERs (type 1 or type 2
Seyferts) are nearly always classified as starbursts (AGN) on the
basis of mid-IR spectroscopy 
%(Lutz et al. 1999; Taniguchi et al. 1999). 
\citep{lu99, ta99}.
Near-IR spectroscopy is also available for much of the
sample, and provides further diagnostic information, such as
identifying AGN on the basis of detection of
broad Pa$\alpha$ or [Si VI] \citep{v99b} and the strength of PAH
emission and absorption features \citep{im00}. Finally, AGN can 
be discriminated from starbursts (on an empirical, statistical basis)
using the $F_{25\mu m}/F_{60\mu m}$ flux ratio, which is a measure of 
dust temperature \citep{de87}. This information is
summarized in Table \ref{proptable}. On the basis of these diagnostics, 5
ULIRGs are 
classified as starbursts  (Arp 220, UGC 5101, IRAS 17208-0014, IRAS
20551-4250, and IRAS 23128-5919), and the other three are classified
as AGN (Mkn 231, Mkn 273, and IRAS 05189-2524).  The analysis of a
3-4$\mu$m spectrum presented by \citet{im01} 
shows that a highly-obscured AGN is
likely to be present in UGC 05101.
In what follows below,
we will refer to these two classes as ``starburst-ULIRGs'' and ``AGN-ULIRGs''
respectively (for consistency we place UGC 05101 in the
starburst-ULIRG category), bearing in mind that {\it this empirical
  classification may 
not reflect the true nature of the dominant energy source.} 
For example, it is fairly clear that an AGN contributes substantially to the
hard X-ray
emission from NGC 6240 \citep{iw98, vi99} even though it is an optical
LINER that appears starburst-dominated in the infrared \citep{lu99,im00}.
We will see
below that the AGN- and starburst-ULIRGs do differ significantly in their hard
X-ray properties. 

%\clearpage
\begin{deluxetable*}{lllllllll}
\tablecaption{ULIRG Parameters and Classification\label{proptable}}
\tabletypesize{\scriptsize}
\tablehead{
\colhead{Galaxy} & \colhead{$L_{FIR}$} & %\colhead{$log L_{[OIII]}$} &
\colhead{$\frac{F_{25\mu m}}{F_{60\mu m}}$} & 
\colhead{$\frac{F_{\rm 2-10\ keV}}{F_{FIR}}$}
& \multicolumn{2}{c}{Classification} & \colhead{$\log
N_{H,CO}$\tablenotemark{*}} & 
\colhead{$R_{CO}$\tablenotemark{$\dagger$}} & \colhead{$R_{2-8 \rm
\ keV}$\tablenotemark{$\ddagger$}} \\
& \colhead{$(10^{12} L_{\odot})$} 
& %\colhead{(ergs s$^{-1}$)} & 
& \colhead{($10^{-4}$)}
&  \colhead{Optical} &
%\colhead{$2\mu m$\tablenotetext{1}} & \colhead{$3-4\mu m$}
\colhead{IR} & \colhead{(cm$^{-2}$)} & \colhead{($\arcsec$)} & 
\colhead{($\arcsec$)}
}
\startdata
Arp 220 & 0.9 & %39.1 & 
0.08 & 0.27 & LINER\tablenotemark{1} &
HII\tablenotemark{2} & 24.7 & 0.9 & 0.9 (0.8-1.1) \\
IRAS 05189 & 0.7 & %41.3 & 
0.25 & 56 & S2\tablenotemark{3} &  AGN\tablenotemark{2},
S1\tablenotemark{4} & & & $<0.28$\\
IRAS 17208 & 1.6 & %39.5 & 
0.05 & 0.44 & HII\tablenotemark{1} & PDR\tablenotemark{5} 
&  24.1 & 1.25 & 1.0 (0.7-1.2) \\
IRAS 20551 & 0.6 & %40.4 &
0.15 & 1.5 & HII\tablenotemark{6} &
HII\tablenotemark{5}%, no broad Pa$\alpha$\tablenotemark{4} 
& & & $<0.70$\\
IRAS 23128 & 0.6 & %41.4 & 
0.15 & 4.8 & HII\tablenotemark{6} &
HII\tablenotemark{5} & & & $<0.47$\\
Mkn 231 & 1.6 & %$<$40.5 & 
0.27 & 8.6  & S1\tablenotemark{3} & 
AGN\tablenotemark{2} & 24.6 & 0.5 & $<0.25$\\ 
Mkn 273 & 0.8 & %41.9 & 
0.11 & 5.8 & S2\tablenotemark{1} &
AGN\tablenotemark{2} & $>$24.6 & $<0.6$ & $<0.47$\\
UGC 05101 & 0.6 & %41.4 & 
0.09 & 1.9 & LINER\tablenotemark{7} &
AGN\tablenotemark{8}, SB\tablenotemark{9} & & & $<0.44$ \\
NGC 6240 & 0.4 & %41.2 & 
0.15 & 17 & LINER\tablenotemark{7} &
SB\tablenotemark{9} & 24.6 & 0.39 & 1.1 (1.0-1.2) % n6240_2-8kev_nucleus_psf_1gauss_1\\
\enddata
\tablecomments{$L_{FIR} = 2.58 L_{60 \micron} + L_{100 \micron}$}
\tablenotetext{*}{Column density (in $\rm cm^{-2}$) inferred from the
CO mass surface brightnesses given in \citet{br99}.}
\tablenotetext{$\dagger$}{Deconvolved semimajor axis of CO emission
given in \citet{br99}.}
\tablenotetext{$\ddagger$}{Deconvolved semimajor axis of 2-10 keV
  emission, based on elliptical Gaussian fits (including PSF model)
  discussed in the text.}
\tablenotetext{1}{\citet{kim98a}}
\tablenotetext{2}{\citet{im00}}
\tablenotetext{3}{\citet{v99a}}
\tablenotetext{4}{\citet{v99b}}
\tablenotetext{5}{\citet{la00}}
\tablenotetext{6}{\citet{ke01}}
\tablenotetext{7}{\citet{v95}}
\tablenotetext{8}{\citet{im01}}
\tablenotetext{9}{\citet{lu99}}
\end{deluxetable*}
%\clearpage

\subsection{Morphology}
In every case, significant hard X-ray emission is present in the
nuclear region.  In general, the hard X-ray nuclei are within
0.5-1.5$\arcsec$ of radio and CO nuclear positions \citep{co91, sc91,
  pl91, ys95, co96, br99, th00}. However, as noted above, in three cases
(the starburst-ULIRGs Arp 220, and IRAS 17208-0014) and in NGC 6240,
 this component is
resolved on arcsecond ($\sim$ kpc) scales \citep[see also ][]{ko03}. 
In the other six cases
where the nuclear source is unresolved, the upper limit to the size
($\sim$ 0.5 kpc FWHM) is consistent with either an AGN or compact
starburst with a size similar to the nuclear radio continuum and CO
emission (see Table \ref{proptable} and references therein).  

In addition to the nuclear emission, a somewhat surprising
result is that often extended hard X-ray emission is detected beyond
the nuclear region, on scales of-order ten kpc. In five cases (the
starburst ULIRGs IRAS 17208-0014, IRAS 20551-4250, IRAS 23128-5919,
UGC 5101) and in NGC 6240, this large-scale emission comprises a significant
fraction ($\sim$ 10 to 50\%) of the total. Thus, in the five
starburst-ULIRGs, the nuclear source is spatially resolved (two cases)
and/or a significant fraction of the total hard X-ray emission arises well
outside the nucleus (four cases). The nuclear source is unresolved and
dominates the total hard X-ray emission in the three AGN-ULIRGs. 

By way of comparison, Chandra observations of
nearby starburst galaxies such as M82 
and NGC 253 have shown that a large fraction of the hard X-ray emission
in these galaxies is due to point-sources, most likely high-mass X-ray
binaries, concentrated within the central $\sim $ kpc of the galaxy.  
Several of the ULIRGs have extra-nuclear sources that would qualify as
ULXs if they are not interlopers.  ULXs are often observed in
starburst galaxies and, more notably for comparison with ULIRGS, in
mergers such as the Antennae \citep{ze02}.
In IRAS 20551-4250 $\sim 50\%$ of the
extra-nuclear hard
X-ray flux may be due to a single ULX.   
The
remainder of the 
hard X-ray emission in starbursts is due to an unresolved population of
lower-luminosity X-ray binaries \citep{gr03}, very hot gas \citep[$\geq$ few
  keV, see][]{ca99,gr00}, 
or
inverse-Compton scattering of IR photons  
off of relativistic electrons generated by supernova shocks
\citep{mo99, pe02}. 
A strong
point source at the dynamical center of the galaxy (i.e. a plausible
AGN candidate) is typically not detected (although cf., Weaver et
al. 2002).

The hard X-ray morphology in these ULIRGs
is generally consistent with these types of starbursts placed at the
same distance as the ULIRG galaxies (even for the AGN-ULIRGs). Table
\ref{proptable} also
lists the semimajor
axis size of nuclear CO observed in several ULIRGs along with the hard
X-ray semimajor axis derived from Gaussian x PSF fits to the hard X-ray
surface brightness.  These values are clearly consistent with the
X-ray emission originating from within the nuclear molecular
cloud/disk, or from a region of similar size (in the cases of Arp 220
and IRAS 17208-0014).

\subsection{Spectroscopy}
% --- This section has been re-written
As stated above, most of the galaxies in this sample require at least two
model components to fit their X-ray spectra.
The soft X-ray spectra of these galaxies show that hot gas with kT $\sim
0.7$ keV is present within the nuclei of these galaxies (with the
possible exception of IRAS 05189-2524 where the AGN may be completely
swamping any thermal flux). In this regard, they are very similar to
typical starburst galaxies observed with Chandra \citep{st02,li02b,ma02}. 
%We will discuss the origin of this hot gas in Paper III.

The hard X-rays we observe may in principle be due to either an AGN or
starburst.  Our analysis shows that the hard X-ray emission detected in
the ULIRGs is generally {\it not} absorbed by large column 
densities (i.e., in excess of $10^{23} \rm \ cm^{-2}$). Mkn 273 is the
only exception \citep[see also ][]{xi02}.  If the hard X-rays are due to an AGN,
then there are two possibilities: the AGN is truly unabsorbed (and hence has
a low intrinsic luminosity) or the AGN is ``Compton-thick'' 
(i.e., absorbed by a
column in excess of $10^{24} \rm \ cm^{-2}$) and the observed hard X-rays are
scattered into the line of sight.  In the first case we would expect the hard
X-ray slopes to be similar to those observed in Seyfert 1 and Compton-thin
Seyfert 2 galaxies, namely $\Gamma = 1.5-2.0$ \citep{mu93, da98, pt99a}.
IRAS 05189-2524 is consistent with this perspective, once we allow for the possible
presence of Fe-K emission with an EW of $\sim 0.1$ keV (see Table 
\ref{fekfits}).  On the
other hand the hard X-ray spectrum of Mkn 231 appears to be genuinely
flat ($\Gamma < 1$).

In the Compton-thick scenario for an AGN origin to the hard X-rays, the hard
X-ray would be due to some combination of reflection from optically-thick
neutral material, scattering from optically-thin (and likely highly-ionized)
material and leakage of X-rays through patches in the obscuring
material. 
These effects would tend to flatten the observed X-ray spectrum and
produce high-EW ($> 1$ keV) Fe-K lines, as observed in samples of Seyfert 2
\citep{tu97} and composite starburst-Seyfert \citep{le01} galaxies
(note that scattering from highly-ionized material alone would not
flatten the observed X-ray spectrum).   The fact
that the 
hard X-ray emission tends to be coincident with CO gas with high implied
column densities (see Table \ref{proptable}) supports this picture
(although a caveat is discussed below), and this
is clearly the case in NGC 6240.  The Fe-K line EW in Mkn 273 is
consistent with that expected to be produced by the transmission of X-rays
through material with column densities of order $10^{23} \rm \ cm^{-2}$
\citep{lh93}, as is observed directly.  However, there are insufficient
counts in most of these spectra to allow 
meaningful constraints to be placed on complex models (i.e., involving
reflection and ``leaky'' absorbers).
Partial-covering fits to the XMM-Newton spectra of IRAS
05189-2524 and UGC 05101 and the Chandra spectrum of Mkn 231
\citep{ga02} are in fact consistent with a highly-obscured hard X-ray
source that is leaky or contains one or more scattering or
reflection regions.  Similar results based on XMM-Newton observations
of ULIRGs are given in \citet{br02}.  Also note that if a highly-ionized
scattering medium is acting as a mirror for the hard 
X-rays, then the Fe-K lines expected from such a mirror would be due to Fe XXV
and Fe XXVI (K$\alpha$ energies of 6.7 and 6.9 keV) with EWs on the
order of 0.5-1.0 keV \citep{bi02}.  
We suggest that in these AGN
ULIRGs the amounts of reflected and/or leaked flux and scattered 
flux are comparable to each other, effectively reducing the equivalent widths
expected from neutral and ionized Fe-K to values within the range of our
upper-limits.  In the case of IRAS 05189-2524, the very tight constraint on
neutral Fe-K suggests that the flux is mostly scattered and not
reflected (or again IRAS 05189-2524 is simply Compton-thin).  

Note that the long-term variability observed in the AGN ULIRGs implies
that both the scattering and reflecting material must be located within
1-2 pc of the central X-ray sources, or distances consistent with the
expected size the putative tori.  However, this argues against the
arcsecond-scale CO gas being a source of reflection that {\it
dominates} the hard X-ray flux in the AGN ULIRGs.  This is consistent
with the relatively high $L_{2-10 \rm \ keV}/L_{IR}$ ratio of AGN
ULIRGS, i.e., the molecular gas is most likely only directly obscuring
the AGN with columns of order $10^{23} \rm cm^{-2}$ or less.  In
general the CO gas appears to be distributed in 100-500 pc scale
disks, and the disks are either viewed face-on and/or (particularly in
the case of double nuclei) the nuclei are not located centrally in the
disks \citep{do98,br99}.  Both of these effects would result in lower
column densities being observed toward the nuclei than the peak values
derived from the CO data.
				   
Turning to the starburst-ULIRGs, the hard X-ray emission
is too weak for strong constraints to be placed on its spectral form
(it is consistent with either a weak AGN viewed through modest
columns, or a typical starburst). Moreover, the upper limits on the Fe
K$\alpha$ EW are not restrictive. 
%If due to a starburst, the hard
%X-ray continuum flux is most likely due to some combination of
%unresolved high mass X-ray binaries
%%(e.g. Grimm et al. 2002),
%\citep{gr03},
%inverse-Compton scattering of IR photons 
%%(e.g. Moran, Lehnert, \&Helfand 1999), 
%\citep{mo99},
%and very hot gas 
%(e.g. Griffiths et al. 2000).  
%\citep{gr00}.
The extended hard X-ray emission resolved by Chandra in NGC 6240 suggests that
starburst processes such as these may be contributing significantly (i.e., the
two nuclei contribute $<$50\% of the 2-10 X-ray flux of NGC 6240 although, as
shown in \citet{ko03}, the nuclei dominate in the 5-8 keV band).
These
possibilities will be discussed in more detail in a subsequent paper.

\subsection{The X-ray Emission of ULIRGs in Context}
\subsubsection{The Far-Infrared}

We have argued above that the signature of an AGN is possibly present
in the hard X-ray emission of at least three of the eight ULIRGs in
our sample (the same three that show evidence for an AGN in their
optical and IR spectra), and that Fe-K emission shows that an
energetically-important AGN is present in two of the starburst
ULIRGs. One way to assess possible energetic 
contribution of AGN is to compare the ratio of the hard X-ray to
far-infrared luminosities in the ULIRGs to values in typical AGN and
starbursts. 

\begin{figure}[htb]
%\plotone{f25_f60_vs_f2-10_fir_19sep02.eps}
\plotone{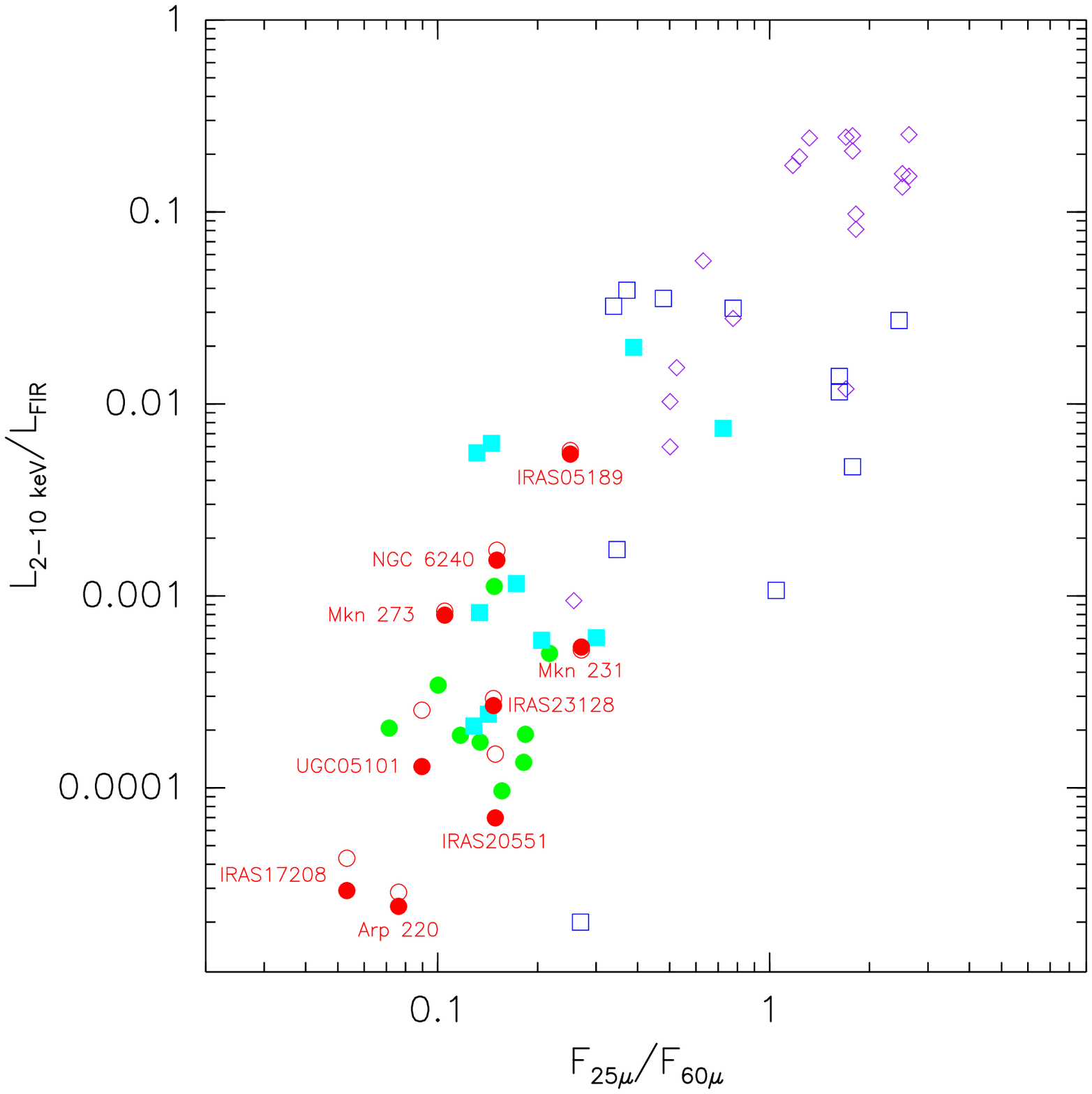}
\caption{$L_{2-10 \ \rm keV}/L_{IR}$ vs. $F_{25\mu}/F_{60\mu}$.
The plot key is: ULIRGS=red circles (values shown with
filled circles are based on X-ray fluxes derived from the nuclear
spectrum), starbursts = green filled
circles, Seyfert 2s = blue squares, composites = cyan filled
squares, Seyfert 1s = purple diamonds.\label{lx_irrat}}
\end{figure}

Following 
%Levenson et al. (2001), 
\citet[hereafter LWH]{le01},
we first plot the ratio of hard X-ray to FIR
luminosity as a function of the $F_{25\mu m}/F_{60\mu m}$ flux ratio
(Figure \ref{lx_irrat}). The latter (a measure of luminosity-weighted mean dust
temperature) is a useful empirical diagnostic to help assess the
relative energetic importance of AGN and 
starburst activity in a galaxy 
%(e.g de Grijp et al. 1987).
\citep{de87}.
  We compare
the ULIRGs to a sample of starbursts observed by ASCA 
\citep{da98, de96, de99, mo99, pt99a, tu97}, a sample 
of composite starburst/Seyfert 2 galaxies, and the "pure" Seyfert 2
and Seyfert 1 galaxies samples from LWH. As shown by
LWH, the pure Seyfert 2's are generally Compton-thin and the
composite starburst/Seyfert 2's are generally Compton-thick.
Since these other samples are based on ASCA data, the 
ULIRG 2-10 keV luminosities were taken from the global spectral fits,
although we also plot the points derived from the nuclear fluxes for
comparison. The IRAS infrared fluxes have been taken from NED, and starburst
galaxies for which the global fluxes fluxes differ by more than
10\% from the point-source values have been taken from \citet{so89}.

%\clearpage
%\clearpage

Several results are clear from this comparison. First, both the mid/far-IR
color and the relative
strength of the hard X-ray continuum emission in the ULIRGs is
generally similar to the values for starburst galaxies. The latter ratio
is typically
about two orders-of-magnitude below the values for most type 1 Seyferts and
pure (Compton-thin) type 2 Seyfert galaxies. IRAS 05189-2524 is the only
ULIRG lying
outside the starburst regime and overlapping the pure type 2 Seyfert
regime. Note that we are using observed hard X-ray luminosities in
these plots, and correcting for the $\sim 10^{23} \rm \ cm^{-2}$
column density in Mkn 273 would increase its 2-10 keV luminosity by a
factor of $\sim 3$ and place it in the HX/FIR $> 10^{-3}$
Compton-thin regime.  There is some overlap between the ULIRGs and the
(Compton-thick) 
composite starburst/Seyfert 2 galaxies,
but the ULIRGs as-a-class are even more extreme.
Thus, in a purely empirical energetic sense there is a need to invoke a
significant contribution of an AGN to the hard X-ray emission in only one
of the eight ULIRGs. Second, the three AGN-ULIRGs (and NGC 6240)
have significantly
higher ratios of hard X-ray to far-IR luminosity than the
starburst-ULIRGs. This suggests that in fact an
AGN {\it is} contributing 
significantly to the hard X-ray emission in these former cases. 

\begin{figure}[htb]
%\plotone{fx_fir_vs_fek_fir_19sep02.eps}
\plotone{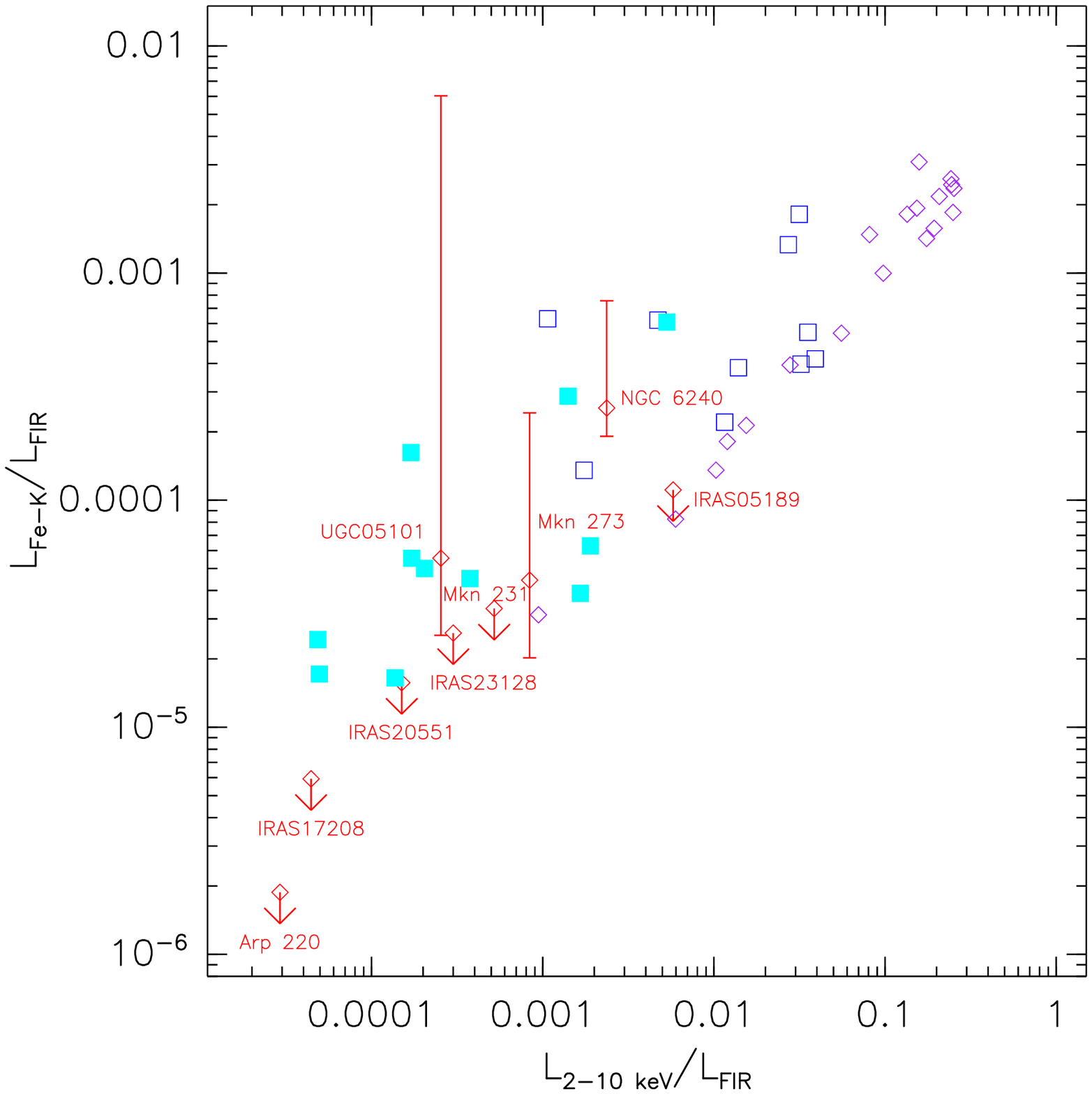}
\caption{$L_{2-10 \ \rm keV}/L_{IR}$ vs. $L_{Fe-K}/L_{IR}$.  
The upper-limits are based on fluxes listed in Table 6.\label{lx_lfek}}
\end{figure}

The above conclusions are reinforced by Figure \ref{lx_lfek}, 
where we have plotted the
ratio of Fe K$\alpha$ and far-IR luminosities {\it vs.} the ratio of hard
X-ray and far-IR luminosities for the above samples of ULIRGs
and AGN. In particular, the upper limits to the ratio of the Fe K$\alpha$
and far-IR fluxes in the ULIRGs lie below even the very low ratios observed
in the Compton-thick starburst/Seyfert 2 composites, and are up to two
orders-of-magnitude lower than in the pure type 2 Seyferts.

%\clearpage
%\clearpage

\subsubsection{The [OIII]$\lambda$5007 Line}

The [OIII]$\lambda$5007 emission-line is the strongest optical line
produced in the kpc-scale Narrow Line Region, and is often used as
a rough indicator of the the true luminosity of the AGN
in both type 1 and type 2 Seyferts. On this basis,
%Basani et al. (1999)
\citet{ba99}
have proposed using the ratio of the hard X-ray
and [OIII]$\lambda$5007 fluxes as an indicator of X-ray absorption 
that is valid even in the Compton-thick regime.
They also show that there is a strong
inverse correlation between the hard X-ray/[OIII] flux ratio and the
Fe K$\alpha$ equivalent width, with Compton-thick type 2 Seyfert nuclei
at one extreme and type 1 Seyferts at the other.

\begin{figure*}[htb]
%\plotone{lOIII_lfek_11sep02.eps}
\plottwo{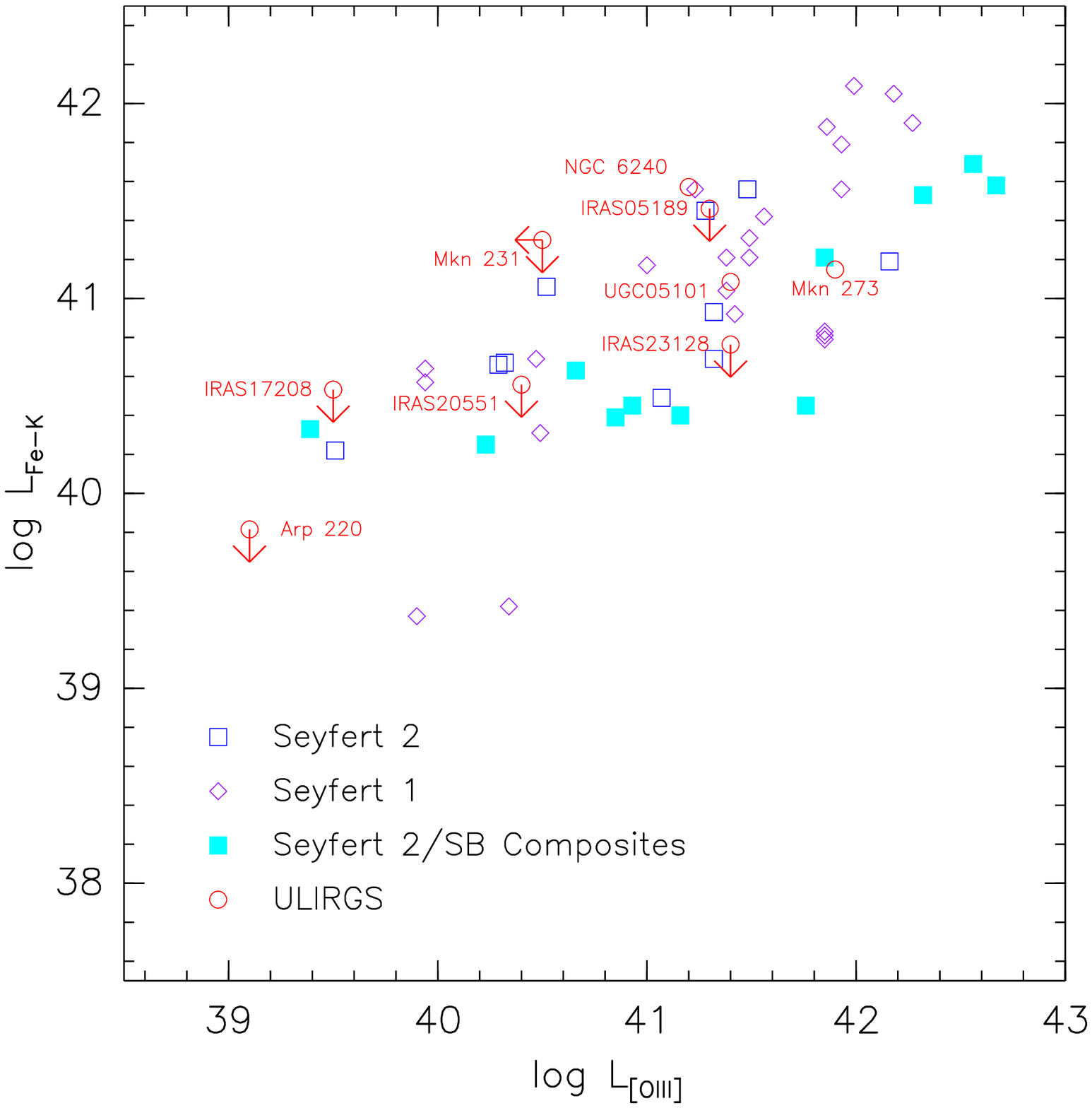}{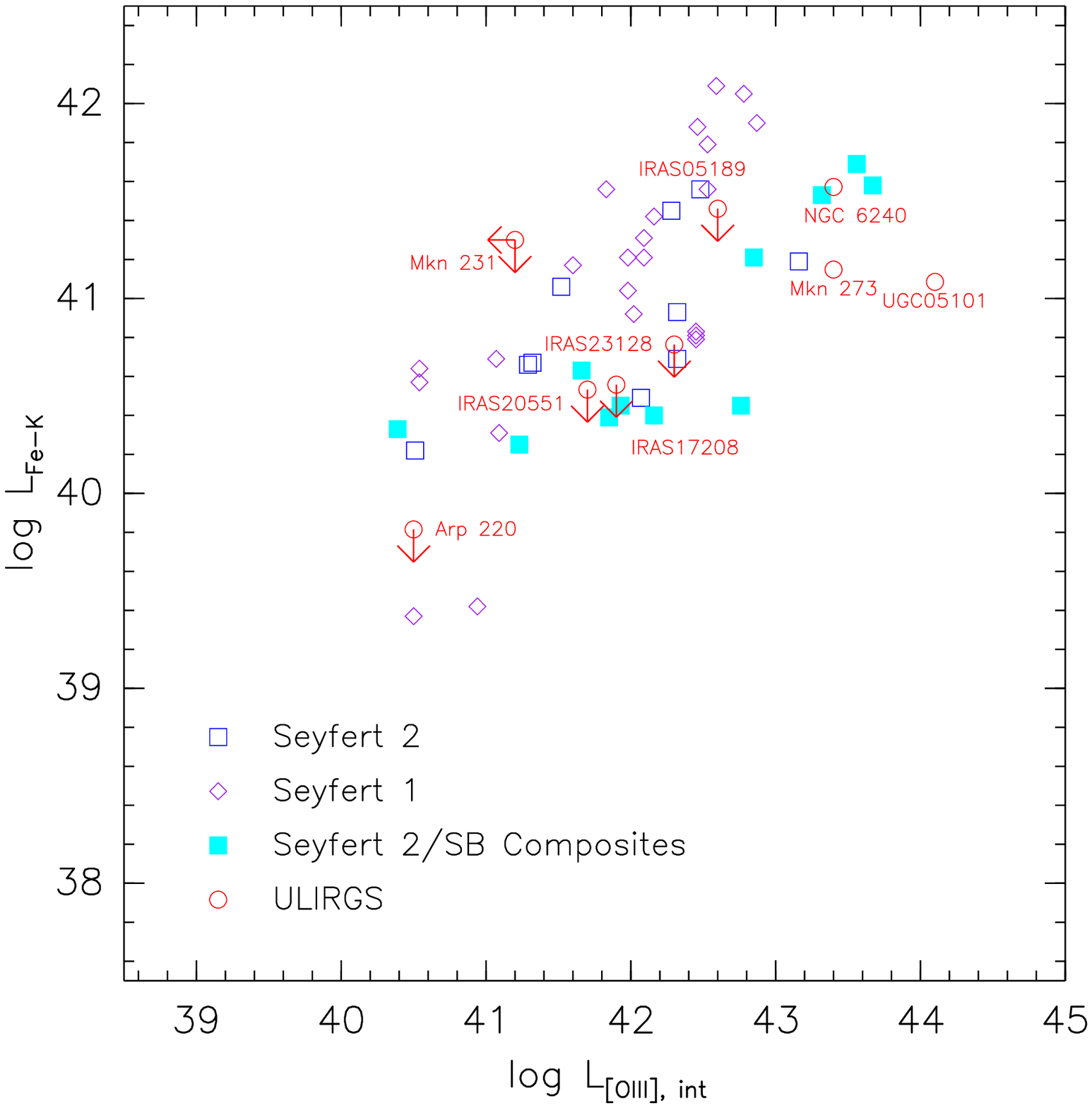}
\caption{Fe-K luminosity
  plotted as a function of [OIII] line luminosity.
(left) Plot generated using observed [OIII]
  luminosities; (right) plot generated using extinction-corrected
  [OIII] luminosities (see text).
The plot key is: ULIRGS=red circles, Seyfert 2s = blue squares,
composites = cyan filled squares, Seyfert 1s = purple
  diamonds.
\label{oiii_fek}
}
\end{figure*}

In Figure \ref{oiii_fek}
we plot the luminosities of the [OIII]$\lambda$5007 lines
{\it vs.} those of the Fe K$\alpha$ lines for Seyferts and ULIRGs.
We have
also corrected the [OIII] for dust extinction using the prescription
(based on the Balmer decrement) given in \citet{da88}. In the case of
the ULIRGs we used the $H\alpha/H\beta$ flux ratio listed in Table
\ref{oiii_data} while we applied mean corrections
\citep[also based on the
statistical analysis in ][]{da88} to the Seyfert 1 and Seyfert 2 samples
(which increased the [OIII] fluxes by factors of 4 and 10,
respectively). In the plot of the extinction-corrected [OIII] luminosity,
the Seyferts exhibit a correlation with roughly unit slope and
a scatter of $\sim\pm$0.5 dex.
This suggests that both the Fe K$\alpha$ line and the [OIII]$\lambda$5007
line can be used as very rough indicators
of the luminosity of the hidden AGN, even in Compton-thick type 2 Seyferts.
Since we have Fe K$\alpha$
detections for only two of the eight ULIRGs, we can only
say that
the data are consistent with the ULIRGs following the same trend as
the Seyferts. More to the point,
the relative weakness of the K$\alpha$ line in the ULIRGs (Figure 
\ref{lx_lfek}) 
suggests that energetically dominant AGN are not present in most
ULIRGs (even though AGN may contribute significantly to the hard
X-ray emission in some cases).

%\clearpage
%\clearpage
% --------------------------------
% 1/2 of the upper-limits in Fig 7b lie above most of the AGN points at
% that [OIII] lum.  Also, to be fair, Sy 2 with only upper-limits on
% Fe-K should also be plotted.

\begin{figure*}[htb]
%\plotone{lOIII_l2-10_20aug02.eps}
\plottwo{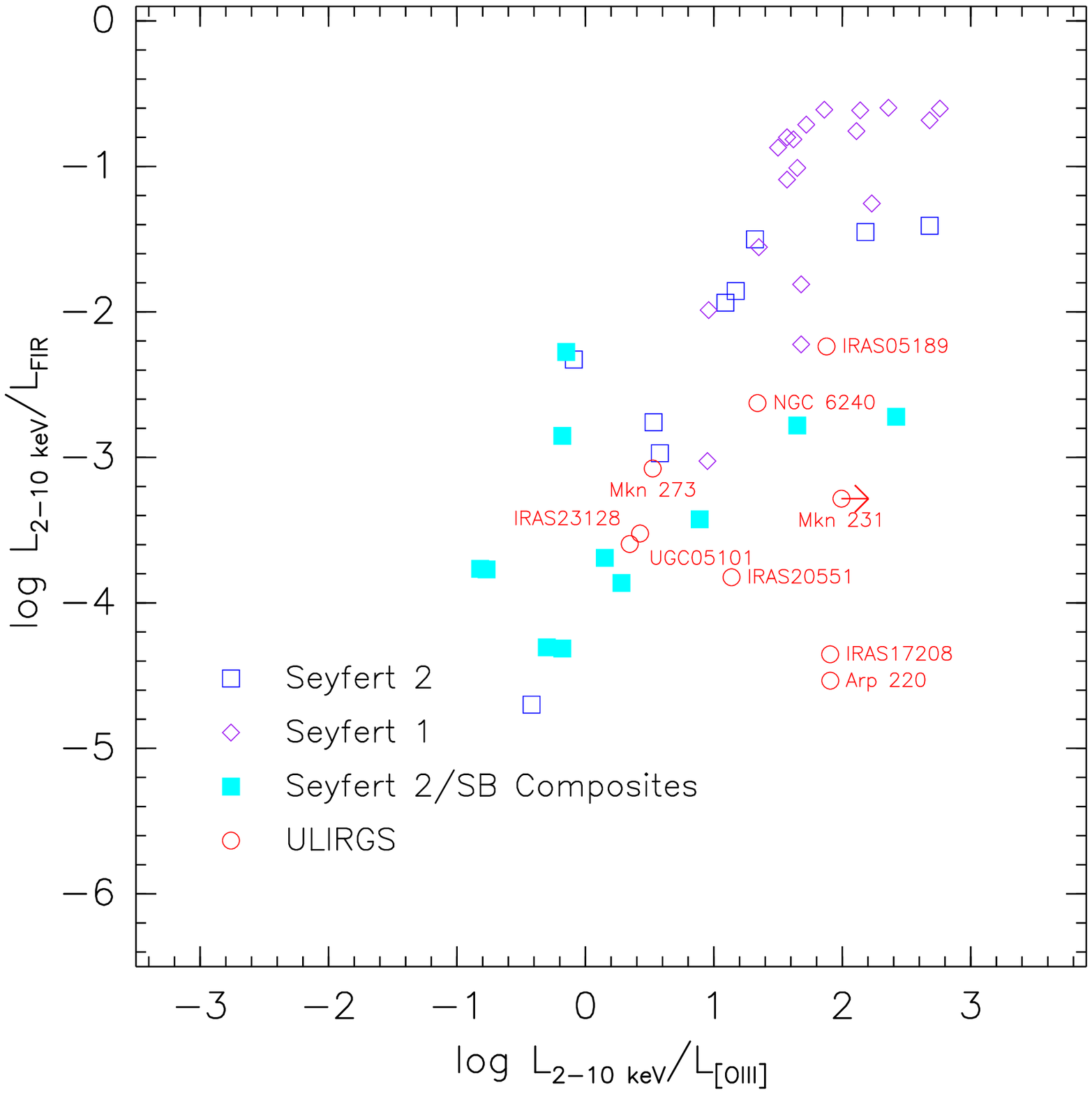}{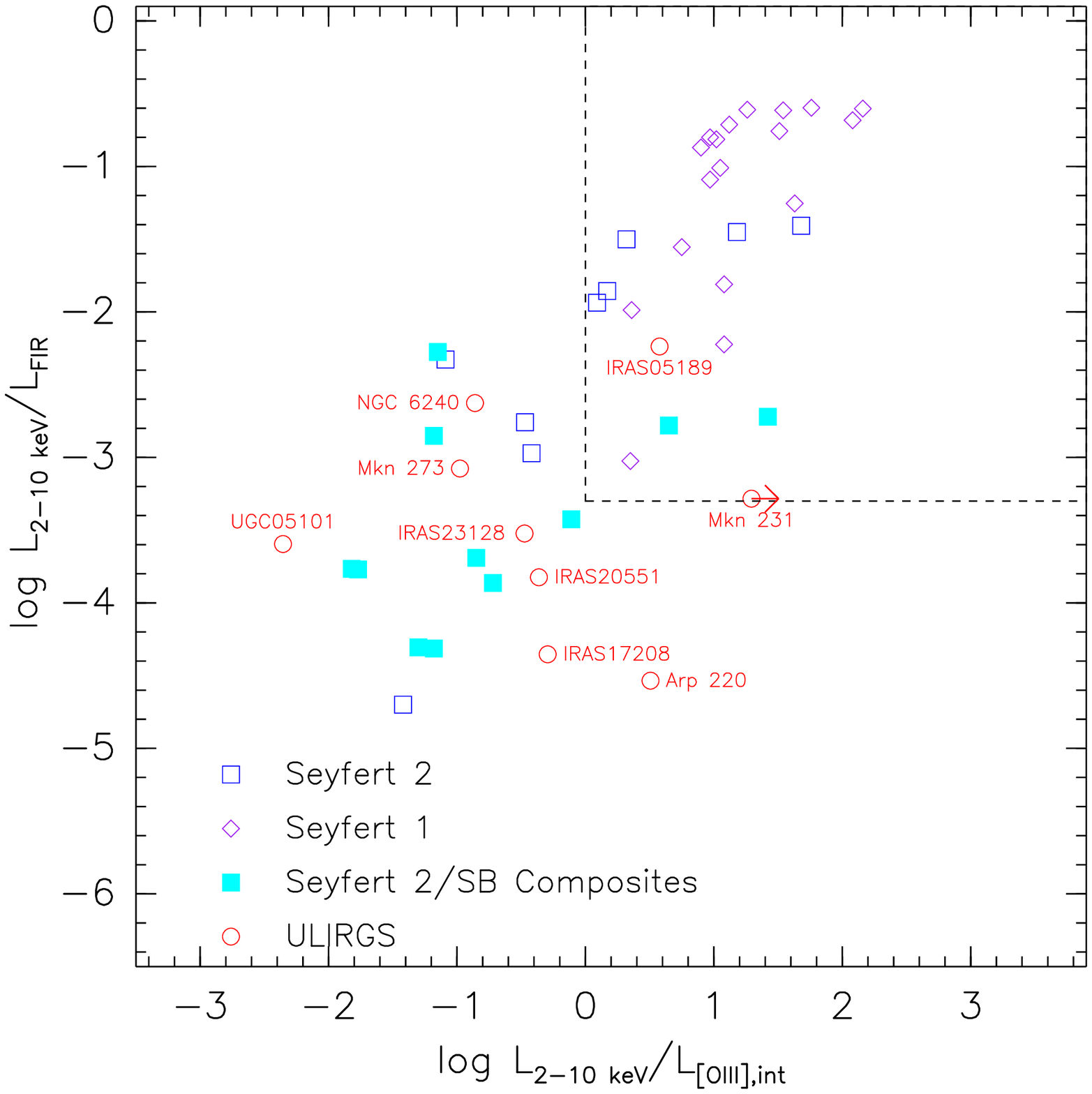}
\caption{2-10 keV/FIR luminosity plotted as a function of 2-10
  keV/[OIII] luminosity. (left) Plot generated using observed [OIII]
  luminosities; (right) plot generated using extinction-corrected
  [OIII] luminosities (see text).  The box delineated in the upper-right
  corner shows the region where Compton-thin Seyfert 2s are found in
  \citet{pa02}. 
The plot key is: ULIRGS=red circles, Seyfert 2s = blue squares, 
composites = cyan filled squares, Seyfert 1s = purple diamonds.
\vspace{2mm}
\label{fx_oiii}
}
\end{figure*}

To further assess the possible contribution of an AGN to the hard
X-ray emission in ULIRGs, we have plotted the ratio of the hard X-ray
and [OIII]$\lambda$5007 flux (see Table \ref{oiii_data}) {\it vs.} the
ratio of the
hard X-ray and far-IR flux
for our samples of ULIRGs and Seyferts (Figure \ref{fx_oiii}). 
Prior to [OIII] extinction correction the
ULIRGs occupy a distinct part of parameter space in this figure. On
the one hand, 
they have ratios of hard X-ray to [OIII] flux that are most similar to type 1
and pure (Compton-thin) type 2 Seyfert galaxies, and are significantly larger
than in the Compton-thick starburst/Seyfert 2 composites. 
Following
\citet{ba99},
this implies only modest X-ray absorption in the ULIRGs. On the other
hand (as discussed above) the ULIRGs have extremely small ratios
of hard X-ray to far-infrared flux. After extinction-correcting the [OIII]
fluxes, the ULIRGs better overlap the locus of
the Compton-thick Seyfert 2 and composite galaxies, albeit
with much more dispersion.  The larger dispersion in the ULIRG sample
is due at least in part to the large (and hence uncertain) corrections
to the [OIII] fluxes.  Taking the extinction-corrected [OIII] fluxes
at face-value, the AGN-ULIRGs Mkn 231 and IRAS 05189-2524 and the
starburst-ULIRG Arp 220 have hard X-ray to [OIII] flux ratios
similar to Compton-thin Seyferts, the AGN-ULIRG Mkn 273, the starburst-ULIRG
UGC 05101, and NGC 6240 have ratios similar to Compton-thick
Seyferts, and the starburst-ULIRGS IRAS 17208-0014, IRAS 20551-4250,
and IRAS 23128-5919 are intermediate.
%\clearpage
\begin{deluxetable*}{lllll}
\tablecaption{ULIRG [OIII] Data\label{oiii_data}}
\tabletypesize{\scriptsize}
\tablehead{
\colhead{Galaxy} & \colhead{$L_[OIII]$} & \colhead{Reference} & 
\colhead{$L_{H\alpha}/L_{H\beta}$} & \colhead{Reference}\\
& \colhead{$(10^{40} \rm \ ergs \ s^{-1}$)}
}
\startdata
Arp 220 & 0.10 & 1 & 9.1 & 1 \\
IRAS 05189-2524 & 62. & 1 & 8.3 & 1 \\
IRAS 17208-0014 & 3.7 & 2 & 18.6 & 3 \\
IRAS 20551-4250 & 28. & 4 & 9.6 & 4 \\
IRAS 23128-5919\tablenotemark{*} & 18. & 4 & 7.0 & 4\\
Mkn 231 & $<$ 3.2 & 5 & 5.4 & 6 \\
Mkn 273 & 18. & 7 & 10. & 3 \\
UGC 05101 & 1.6 & 8,9 & 23.4 & 8\\
NGC 6240 & 16. & 10, 11 & 16.1 & 8 \\
\enddata
\tablerefs{
1. \citet{v99a}; 2. \citet{kim98a}; 3. \citet{v99b}; 4. \citet{du97};
5. unpublished data; 6. \citet{da88}; 7. \citet{wi92}; 8. \citet{v95};
9. \citet{ki95}; 10. \citet{ar89}; 11. \citet{ar90}
}
\tablenotetext{*}{Values cited are for the southern nucleus of IRAS
  23128-5919, which dominates the hard X-ray emission.}
\end{deluxetable*}

\section{Summary}
We have presented the initial results from a Chandra survey of a
complete sample of the 8 nearest ($z \leq$ 0.04) ultraluminous IR
galaxies (ULIRGs) plus NGC 6240, using the  hard X-rays (2-8 kev) to
search for the 
possible presence of an obscured AGN, and to assess its contribution
to the bolometric luminosity. In six cases (including NGC 6240), the
extant optical and 
infrared spectra suggest that a starburst dominates the energetics
(the "starburst-ULIRGs"), while an energetically significant AGN is
present in the other three (the "AGN-ULIRGs"). We find that the hard
X-ray properties of these two subtypes differ as well. 

A hard X-ray source is detected in the nuclear region in every
case. The nuclear source is spatially-resolved in two of the
starburst-ULIRGs Arp 220 and IRAS 17208-0014 (FWHM $\sim$ 1 kpc and 4
kpc respectively) and also in NGC 6240, and is unresolved (FWHM
$\lesssim$ 0.5 kpc) in the 
others. The upper limits to the spatial extent in the six unresolved
cases are consistent with either an AGN or a compact starburst. Hard
X-ray emission on 
larger (galactic) spatial scales is significant in six cases (all
starburst-ULIRGs), comprising 10 to 50\% of the total flux. We have
shown that five starburst-ULIRGs have hard X-ray luminosities about
an order-of-magnitude smaller than the three AGN-ULIRGs, with the hard X-ray
luminosity of NGC 6240 being comparable to the AGN-ULIRGs. 

Our analysis of the hard X-ray spectra provides no direct evidence for
absorbing columns in excess of $\sim10^{23}$ cm$^{-2}$ except for in
Mkn 273, although the
data are relatively poor for the five starburst-ULIRGs.
The Fe K$\alpha$ line is convincingly detected in Mkn 273 and NGC 6240
and marginally detected in UGC 05101. 
% --- Commented out 10/1/02
%The measured values and typical upper 
%limits ($\sim$ keV) for the K$\alpha$ equivalent width are consistent
%with X-ray transmission through moderate column densities ($\leq
%10^{24}$ cm$^{-2}$). 
%Corroboration
%of the modest absorbing columns is provided by the ratios of the
%hard X-ray and [OIII]$\lambda$5007 fluxes in the ULIRGs, which are similar to
%the ratios in Compton-thin Seyferts and significantly larger than
%in Compton-thick type 2 Seyferts. 
%This suggest that the hard
%X-ray emission in ULIRGs is intrinsically weak, rather than
%intrinsically strong but heavily absorbed, with NGC 6240 being a
%prominent exception.

The ratio of the hard X-ray to far-IR flux in the ULIRGs (HX/FIR) is
about three (two) orders of magnitude smaller than in type 1 
(Compton-thin type 2)
Seyfert galaxies. Only the AGN-ULIRG IRAS05189-2524 has a value for
HX/FIR that is significantly higher than that in typical starburst
galaxies, and only IRAS 05189-2524 and Mkn 231 are in the Compton-thin regime
of a 
HX/FIR {\it vs.} HX/[OIII] plot. The three X-ray-brightest ULIRGS (the
AGN-ULIRGs) and NGC 
6240 have
intermediate values of HX/FIR similar to the Seyfert 2/ starburst
composite systems studied by LWH.  Likewise, the
flux ratio of the Fe K$\alpha$ line and far-IR continuum is usually at
least two orders-of-magnitude smaller in the ULIRGs than in typical
Seyfert galaxies (but with some overlap with the Seyfert2/starburst
composites). We show that type 1 and type 2 Seyferts follow a 
%rather tight 
correlation between the [OIII]$\lambda$5007 and Fe K$\alpha$
luminosities. The weakness of the [OIII] and Fe K$\alpha$ emission in
the ULIRGs (relative to the far-IR) suggests a correspondingly weak
AGN, although a highly-absorbed AGN (possibly) in conjunction with a complex
scattering geometry cannot be ruled out.

Existing optical and IR spectroscopy show that AGN are definitely
present in three of the ULIRGs, and we conclude that these AGN make a
significant contribution to the observed hard X-ray emission in these
cases. There is no compelling reason to
invoke the presence of an AGN 
in the other five cases except for the strong Fe-K emission in UGC
05101 and NGC 6240 (where the AGN contribution to hard X-rays is also
well established from the BeppoSAX PDS data; Vignati et al. 1999). 
Even in the three AGN-ULIRGs, the
contribution of the AGN to the bolometric (IR) luminosity is quite
uncertain. To make a major contribution, an AGN must be buried behind
highly Compton-thick material with a very small transmitted or
reflected fraction compared to typical Seyfert galaxies.
This could be possible (given
the very large column densities of molecular gas observed in the
nuclei of these galaxies), but our new data provide no direct evidence
for such high absorbing columns. 

Of course, absence of evidence is not evidence of absence: it is
exceedingly difficult to robustly prove that powerful AGN can not
be present in these galaxies. However, we 
conclude that our new data
provide no evidence that powerful ``buried quasars'' dominate the
overall energetics of most ULIRGs, particularly those with a starburst
optical or IR classification. 

% -------------------
% Again scattering into the line of sight from ionized gas might be
% common, or a very buried agn with starburst producing the hard X-ray
% flux, both of which would result in low or no observed Fe-K EW.   Case in
% point is the fact that the ulirgs are fairly consistent with the
% composites and some Sy2/NELG in figure 5 and 7a and the L_Fe-K limits are
% generally above the composite Fe-K levels in 7b.

\acknowledgments
We would like to thank the anonymous referee for useful comments.
This work made use of the NASA Extragalactic Database (NED), the NASA
High-Energy Astrophysics Science and Research Center (HEASARC), the
NASA Astrophysics Data System Bibliographic Services, and the Chandra
X-ray Center Chandra Data Archive. This work was supported by the NASA
grants NAG5 9910 and GO1-2086X.


\begin{thebibliography}{}
\bibitem[Armus, Heckman \& Miley(1989)]{ar89}Armus, L., Heckman, T.,
  \& Miley, G. 1989, \apj, 347, 727
\bibitem[Armus, Heckman \& Miley(1990)]{ar90}Armus, L., Heckman, T.,
  \& Miley, G. 1990, \apj, 364, 471
\bibitem[Arribas, Colina, \& Clements(2001)]{ar01}Arribas, S., Colina,
L., \& Clements, D. 2001, \apj, 560, 160
\bibitem[Bassani et al.(1999)]{ba99}Bassani, L. et al. 1999, \apjs, 121, 473
\bibitem[Bianchi \& Matt(2002)]{bi02}Bianchi, S. \& Matt, G. 2002,
  \aa, 387, 76
\bibitem[Blain et al.(1999)]{bl99}Blain, A., Smail, I., Ivison, R., \&
Kneib, J.-P. 1999, \mnras, 302, 632 
\bibitem[Braito et al.(2002)]{br02}Braito, V. et al. 2002,
Proc. Symnposium ``New Visions of the X-ray Universe in the XMM-Newton
and Chandra Era", astro-ph/0202352
\bibitem[Bryant \& Scoville(1999)]{br99}Bryant, P. \& Scoville,
N. 1999, \apj, 117, 2632
\bibitem[Cappi et al.(1999)]{ca99}Cappi, M. et al. 1999, \aa, 350, 777
\bibitem[Cash(1979)]{ca79}Cash, W. 1979, \apj, 228, 939
\bibitem[Clements, Sutherland, McMahon, \& Saunders
(1996)]{cl96}Clements, D., Sutherland, W., McMahon, R. \& Saunders,
W. 1996, \mnras, 279, 477
\bibitem[Clements et al.(2002)]{cl02}Clements, D. et al. 2002, \apj, 581, 974
\bibitem[Condon et al.(1991)]{co91}Condon, J., Huang, Z., Yin, Q., \&
Thuan, T. 1991, \apj, 65
\bibitem[Condon et al.(1996)]{co96}Condon, J., Helou, G., Sanders,
D., \& Soifer, B. 1996, \apjs, 103, 81
\bibitem[Dahari \& De Roberts(1988)]{da88}Dahari, O. \& De Roberts,
  M. 1988, \apjs, 67, 249
\bibitem[Dahlem, Weaver \& Heckman(1998)]{da98}Dahlem, M., Weaver,
K. \& Heckman, T. 1998, \apjs, 118, 401
\bibitem[Davis(2001)]{da01}Davis, J. 2001, \apj, 562, 575
\bibitem[de Grijp,  Miley, \& Lub(1987)]{de87}de Grijp, M., Miley,
  G., \& Lub, J. 1987, \aaps, 70, 95
\bibitem[Duc, Mirabel, \& Maza(1997)]{du97}Duc, P.-A., Mirabel, I. \&
Maza, J. 1997, \aap, 124, 533
\bibitem[Della Ceca et al.(1996)]{de96}Della Ceca, R., Griffiths, R.,
Heckman, T., \& MacKenty, J. 1996, \apj, 469, 662
\bibitem[Della Ceca et al.(1999)]{de99}Della Ceca, R., Griffiths, R.,
Heckman, T., Lehnert, M., \& Weaver, K. 1999, \apj, 514, 772
\bibitem[Dickey \& Lockman(1990)]{di90}Dickey, J. \& Lockman, F. 
1990, Ann. Rev. Ast. Astr. 28, 215
\bibitem[Downes \& Solomon(1998)]{do98}Downes, D. \& Solomon, P. 1998,
  \apj, 507, 615
\bibitem[Eracleous et al.(2002)]{er02}Eracleous, M., Shields, J., Chartas,
  G., \& Moran, E. 2002, \apj, 565, 108
\bibitem[Ferrarese \& Merritt(2000)]{fe00}Ferrarese, L. \& Merritt, D.
2000, \apj, 539, L9
\bibitem[Gallagher et al.(2002)]{ga02}Gallagher, S., Brandt, W.,
Chartas, G., Garmire, G., \& Sambruna, R. 2002, \apj, 569, 655
\bibitem[Genzel et al.(1998)]{ge98}Genzel, R., Lutz, D., Sturm, E., Egami,
E., Kunze, D., Moorwood, A., Rigopoulou, D., Spoon, H., Sternberg, A., 
Tacconi-Garman, L. Tacconi, L., \& Thatte, N. 1998, \apj, 498, 579
\bibitem[Genzel et al.(2001)]{ge01}Genzel, R., Tacconi, L., Rigopoulou, D.,
Lutz, D., \& Tecza, M. 2001, \apj, 563, 527
\bibitem[Griffiths et al.(2000)]{gr00}Griffiths, R., et al. 2000, Science,
290, 1325
\bibitem[Grimm, Gilfanov, \& Sunyaev(2003)]{gr03}Grimm, H., Gilfanov, M.,
\& Sunyaev, R. 2003, \mnras, 339, 793
%submitted(astro-ph/0205371)
%\bibitem[Heckman et al.(2002)]{he02}Heckman, T. et al. 2002, in prep.
\bibitem[Imanishi \& Dudley(2000)]{im00}Imanishi, M. \& Dudley,
C.2000, \apj, 545, 701
\bibitem[Imanishi, Dudley \& Maloney(2001)]{im01}Imanishi, M.,
Dudley, C., \& Maloney, P. 2001, \apjl, 558, 93
\bibitem[Iwasawa \& Comastri(1998)]{iw98}Iwasawa, K. \& Comastri,
A. 1998, \mnras, 297, 1219
\bibitem[Iwasawa(1999)]{iw99}Iwasawa, K. 1999, \mnras, 302, 96
\bibitem[Joseph(1999)]{jo99}Joseph, R. 1999, \apss, 266, 321
\bibitem[Kewley, Heisler, Dopita, \& Lumsden(2001)]{ke01}
Kewley, L., Heisler, C., Dopita, M., \& Lumsden, S. 2001,
\apjs, 132, 37
\bibitem[Kim, Veilleux, \& Sanders(1995)]{ki95}Kim, D.-C., Veilleux,
S., \& Sanders, D. 1998, \apjs, 98, 129
\bibitem[Kim, Veilleux, \& Sanders(1998)]{kim98a}Kim, D.-C., Veilleux,
S., \& Sanders, D. 1998, \apj, 508, 627
\bibitem[Kim \& Sanders(1998)]{kim98b}Kim, D.-C. \& Sanders, D. 1998,
\apjs, 119, 41
\bibitem[Komossa et al.(2003)]{ko03}Komossa, S., Burwitz, V.,
Hasinger, G., Predehl, P., Kaastra, J. S., \& Ikebe, Y. 2003, \apj,
582, L15
\bibitem[Krolik \& Kallman(1987)]{kk87}Krolik, J. \& Kallman,
T. 1987, \apj, 320, 5
\bibitem[Laurent et al.(2000)]{la00}Laurent, O., Mirabel, I.,
Charmandaris, V., Gallais, P., Madden, M., Vigroux, L., \& Cesarsky,
C. 2000, \aap, 359, 887
\bibitem[Levenson, Weaver, \& Heckman(2001)]{le01}Levenson, N. A.,
Weaver, K., \& Heckman, T. 2001, \apj, 
550, 230
\bibitem[Levenson et al.(2002)]{le02}Levenson, N., Krolik, J., Zycki, P.,
  Heckman, T., Weaver, K., Awaki, H., \& Terashima, Y. 2002, \apj, 573, 81L
\bibitem[Leahy \& Creighton(1993)]{lh93}Leahy, D. \& Creighton,
  J. 1993, \mnras, 263, 314
\bibitem[Liedahl, Osterheld, \& Goldstein(1995)]{li95}Liedahl, D.A.,
  Osterheld, A.L., and Goldstein, W.H. 1995, ApJL, 438, 115
\bibitem[Lira et al.(2002)]{li02a}Lira, P., Ward, M., Zezas, A.,
  Alonso-Herrero, A., \& Ueno, S. 2002, \mnras, 330, 259
\bibitem[Lira, Ward, Zezas \& Murray(2002)]{li02b}Lira, P., Ward, M.,
Zezas, A. \& Murray, S. 2002, \mnras, 333, 709
\bibitem[Lutz, Veilleux, \& Genzel(1999)]{lu99}Lutz, D., Veilleux,
S., \& Genzel, R. 1999, \apj, 517, 13
\bibitem[Magdziarz \& Zgziarski(1995)]{ma95}Magdziarz, P. \&
  Zgziarski, A. 1995, \mnras, 273, 837
\bibitem[Martin, Kobulnicky, \& Heckman(2002)]{ma02}Martin, C.,
  Kobulnicky, H., \& Heckman, T. 2002, \apj, 574, 663
\bibitem[McDowell et al.(2003)]{mc03}McDowell, J. et al. 2003, \apj, in press, astro-ph/0303316
\bibitem[Misaki et al.(1999)]{mi99}Misaki, K., Iwasawa, K.,
Taniguchi, Y., Terashima, Y., Kunieda, H., Watarai, H. 1999, Adv. Sp. Research,
23, 1051
\bibitem[Moran, Lehnert, \& Helfand(1999)]{mo99}Moran, E., Lehnert,
M., \& Helfand, D. 1999, \apj, 526, 649
\bibitem[Mushotzky, Done, \& Pounds(1993)]{mu93}Mushtozky, R., Done, C., \&
  Pounds, K. 1993,  \araa, 31, 717
\bibitem[Norman \& Scoville(1988)]{ns88}Norman, C. \& Scoville,
N. 1988, \apj, 332, 124
\bibitem[Norris(1988)]{no88}Norris, R. 1988, \mnras, 230, 345
\bibitem[Panessa \& Bassani (2002)]{pa02}Panessa, E. \& Bassani, L. 2002, \aa, 394, 435
\bibitem[Persic \& Rephaeli(2002)]{pe02}Persic, M., \& Rephaeli,
  Y. 2002, \aa, 382, 843
\bibitem[Planesas, Mirabel, \& Sanders(1991)]{pl91}Planesas, P.,
Mirabel, I., \& Sanders, D. 1991, \apj, 370, 172
\bibitem[Ptak et al.(1996)]{p96}Ptak, A., Yaqoob, T., Serlemitsos,
P. J., Kunieda, H., \& Terashima, Y. 1996, \apj, 459, 542
\bibitem[Ptak et al.(1999)]{pt99a} Ptak, A, Serlemitsos, P., Yaqoob,
T., Mushotzky, R. 1999, \apjs, 120, 179
\bibitem[Ptak \& Griffiths(1999)]{pt99b}Ptak, A. \& Griffiths,
R. 1999, \apjl, 517, 85
\bibitem[Ptak \& Griffiths(2002)]{pt02}Ptak, A. \& Griffiths, R. 2002,
  Astronomical Data Analysis Software and 
Systems XII, ASP Conference Proceedings Series, in press., astro-ph/0303104
%\bibitem[Ptak et al.(2002)]{pt02}Ptak, A. et al. in prep.
%\bibitem[Ranalli, Comastri, \& Setti(2002)]{ra02}Ranalli, P.,
%Comatri, A., \& Setti, G. 2002, Proceedings of the Symposium "New
%Visions of the X-ray Universe in the XMM-Newton and Chandra Era",
%astro-ph/0202241
\bibitem[Risaliti, Gilli, Maiolino, \& Salvati(2000)]{ri00}Risaliti,
G., Gilli, R., Maiolino, R. \& Salvati, M. 2000, \aap, 357, 13
\bibitem[Rothernflug \& Arnaud(1985)]{ra85}Rothernflug, R. \& Arnaud,
M. 1985, \aap, 144, 431
\bibitem[Sanders \& Mirabel(1996)]{sa96}Sanders, D. \& Mirabel,
I. 1996, \araa, 34, 749
\bibitem[Sanders(1999)]{sa99}Sanders, D. 1999, \apss, 266, 331
\bibitem[Scoville et al.(1991)]{sc91}Scoville, N., Sargent, A.,
Sanders, D., \& Soifer, B. 1991, \apj, 366, 5
\bibitem[Severgnini et al.(2001)]{se01}Severgini, P, Risaliti, G.,
Marconi, R., \& Salvati, M. 2001, \aap, 368, 44
\bibitem[Soifer et al.(1989)]{so89}Soifer, B., Boehmer, G.,
Neugebauer, G., \& Sanders, D. 1989, \aj, 98, 766
\bibitem[Strickland et al.(2002)]{st02}Strickland, D., Heckman, T.,
  Weaver, K., Hoopes, C., \& Dahlem, M. 2002, \apj, 568, 689
\bibitem[Taniguchi, Yoshino, Ohyama, \& Nishiura(1999)]{ta99}
Taniguchi, Y., Yoshino, A., Ohyama, Y., \& Nishiura, S. 1999, \apj,
514, 660
\bibitem[Terashima et al.(2002)]{te02}Terashima, Y., Iyomoto, N., Ho,
L., \& Ptak, A. 2002, \apjs, 139, 1
\bibitem[Terashima, Ho, \& Ptak(2000)]{te00}Terashima, Y., Ho, L., \& Ptak, A. 2000, ApJ, 539, 161
\bibitem[Thean et al.(2000)]{th00}Thean, A., Pedlar, A., Kukula, M., Baum, S. \& O'Dea, C. 2000, \mnras, 314, 573
\bibitem[Tremaine et al.(2002)]{tr02}Tremaine, S., Gebhardt, K.,  Bender, R., Bower, G., Dressler, A.,
Faber, S., Filippenko, A., Green, R., Grillmair, C., Ho, L., Kormendy, J.,
Lauer, T., Magorrian, J., Pinkney, J., \& Richstone, D. 2002, \apj, 574, 740
\bibitem[Turner et al.(1997)]{tu97}Turner, T., George, I., Nandra,
K., \& Mushotzky, R. 1997, \apjs, 113, 23
\bibitem[Yun \& Scoville(1995)]{ys95}Yun, M. \& Scovilla, N. 1995,
\apj, 451, L45
\bibitem[Veilleux, Kim, Sanders, Mazzarella, \& Soifer
(1995)]{v95}Veilleux, S., Kim, D.-C., Sanders, D., Mazzarella, J., \&
Soiffer, B. 1995, \apjs, 98, 171
\bibitem[Veilleux, Kim, \& Sanders(1999)]{v99a}Veilleux, S,
Kim, D.-C., \& Sanders, D.  1999, \apj, 522, 113
\bibitem[Veilleux, Sanders, \& Kim(1999)]{v99b}Veilleux, S,
Sanders, D.,\& Kim, D.-C. 1999, \apj, 522, 139
\bibitem[Vignati et al.(1999)]{vi99}Vignati, P. et al. 1999, \aa,
  349, L57
\bibitem[Weaver, Heckman, \& Dahlem(2000)]{we00}Weaver, K., Heckman,
T. \& Dahlem, M. 2000, \apj, 534, 684
\bibitem[Weaver et al.(2002)]{we02}Weaver, K., Heckman, T., Strickland, D.,
  \& Dahlem, M. 2002, \apj, 576, 19L
\bibitem[Wittle (1992)]{wi92}Wittle, M. 1992, \apjs, 79, 49
\bibitem[Xia et al.(2002)]{xi02} Xia, X., Xue, S., Mao, S., Boller,
T., Deng, Z., \& Wu, H. 2002, \apj, 564, 196
\bibitem[Zezas \& Fabbiano(2002)]{ze02}Zezas, A. \& Fabbiano, G. 2002,
  \apj, 577, 726
\end{thebibliography}
\end{document}